\nonstopmode
\documentstyle[aps,eqsecnum]{revtex}
\input boxedeps 
\input epsf
\begin{document} 
\draft 
\title{Crescent singularities and stress focusing in a developable cone}
\author{Sahraoui Cha\"{\i}eb and Francisco Melo}
 \address{Departamento de F\'{\i}sica de la Universidad de Santiago, Avenida  Ecuador 
 3493, Casilla 307, Correo 2, Santiago, Chile} 
\vskip .2cm
\author{\parbox{430pt}{\vglue 0.3 cm  {In this paper we investigate
developable cones ({\em d}-cones) topology and mechanical properties. 
We found that for a  sample of a finite 
thickness the singularity is never pointlike but has a spatial extension in form of a crescent. 
The variations of the sheet local curvature versus the distance from the 
{\em d}-cone tip for all deformations showed that the {\em d}-cone tip is rejected by a 
distance which depends on the deformation and on the singularity extension.
A further deformation of the {\em d}-cone leads to  a transition to a plastic 
deformation equivalent to a decrease in the singularity size characterized from 
curvature and profile analysis. The crescent radius of curvature is measured
both at small deformations and at large deformations. It is found that during the 
buckling process, the curvature of the crescent exhibit two different scaling versus the
deformation. From the cone
profile, we measured the reaction force of the plate to deformation and from 
force measurements, the energy necessary to create the singularity is characterized. } 
\vskip 0.1in
\pacs{03.40.Dz, 46.30.-i, 62.20.Fe, 68.55.Jk}}}

\maketitle
\vskip .8mm
\protect
\section{Introduction}
\label{intro}
Strong deformations of membranes and thin shells span a wide range of scales. 
On the microscopic scale, quenched disorder in partially polymerized membranes 
and thermal fluctuations induce, without strain, a crumpling transition at the 
melting point, below which the membrane behaves like a 2D solid. 
At the crumpling transition, partially polymerized 
vesicles look like dried prunes \cite{ben,kantor,chaieb}. 
Some inorganic compounds such as nanotubes were observed in a phase that look 
similar to a crumpled sheet \cite{spec} where they could be buckled like macroscopic
sheets \cite{falvo}.   
In much larger scales, in (2+1)-dimensional general 
relativity, defects-induced deformations of  a two dimensional sheet is characterized 
by a presence of conical singularities\cite{deser}. As Einstein's equations 
(similar to the F\"oppl-von K\'arm\'an (FvK) ones) are of a higher order in the derivation 
than the integrand of the functional from which minimization they are derived, 
they could leave some room for the occurrence of singularities like the tip of a 
developable cone (which are surfaces with a shape of a cone but 
obtained from a plane isometrically) introduced in \cite{pome}. It could 
be possible that the notion 
of developable cones has  applications to general relativity. A developable surface,
is known as a surface that can be obtained from, or applied to, a plane without 
changing distances. Unlike developable surfaces, a developable cone has a zero gaussian 
curvature everywhere except at the tip, called the {\em singularity} where the 
curvature diverges. In intermediate scales, 
stability of shells and thin elastic plates is of a great importance 
in structure engineering and packaging material development \cite{lobko}. When
a thin elastic sheet is confined to a region much smaller than its size the morphology
of the resulting crumpled membrane is a network of straight ridges or folds that
meet at sharp points or vertices. 
Singularities that appear on such a  crumpled sheet, 
as a result of the stress focusing, have been recently the subject of several 
investigations\cite{lobko,amar,witt,kramer,lobko1,lobko2,chaieb1,chaieb2,maha}. 
For instance, in the 
case of a crumpled sheet, {\em d}-cones  were found to be the
solution to FvK equations for large deflections \cite{amar}.  
A scaling analysis of the FvK  equations showed that  strain and 
deformation energy are located within the ridge region that separates 
two singularities\cite{lobko1}. In practice, it was shown that singularity 
energy plays an important role in selecting characteristic lengths in a 
axially buckled cylindrical sheet; These lengths were shown to be the distance 
separating two {\em d}-cones \cite{chaieb1} and whose selection was due to a 
competition between bending energy which favors large 
creases  by flattening the surface and the singularity energy which favors smaller 
creases by respecting geometrical constraints like the natural curvature of the cylindrical 
shell.  
This linear relation between the crease length and the panel radius found experimentally
in \cite{chaieb1} was imposed, between the crease length and the radius of a ball
made by crumpling an elastic sheet, as a condition to fulfill the scaling of the deformation 
energy versus the crease length \cite{kramer}. In a study of a conical singularity, the 
shape od a developable cone was calculated from the condition of zero
in-plane stress and developability \cite{maha}. However a study of the postbuckling state
is still lacking which could explain the appearence of the irreversible crescent shape 
of conical singularities in a crumpled sheet \cite{chaieb3}. In the following we will
show that the behavior of the crescent is associated with the plastic transition.

In this paper we study mechanical and topological properties of the crescent singularity
left after post-buckling a circular sheet of thickness $h$. Unlike zero thickness sheets studied 
theoretically, the singularity in a real sheet is not a pointlike vertex but has a spatial
extension over a radius $R_c$. This crescent appears as a strain-localization induced
curvature focusing at the ridge separating the convex region and the concave region of 
the {\em d-cone}. This focusing is tested by measuring the growth of the curvature  
on the ridge and on the concave part. It is also revealed by measuring the reaction 
force of the plate at the ridge and at the concave part. The {\em singularity energy } is 
measured as the energy dissipated when the crescent appears.

This paper begins with a description of the setup. In Sec. III we present the 
profiles of the d-cone obtained, from which we retrieve the opening angle and 
the aperture angle. In Sec. IV we discuss a simple model that describes the properties 
of the d-cone as an isometric
deformation obtained by pure bending. In Sec. V, we present and discuss the local 
properties, topological and mechanical, of the d-cone. In Sec. VI we present force 
measurements from profiles and from direct load measurements, this latter 
allows us to measure the singularity energy. Finally, an eventual analogy between 
the d-cone and the dislocation problem is discussed and future implications are 
presented in Sec. VII.
\protect
\section{Experimental setup}

\begin{figure}
\centerline{\epsfxsize=\columnwidth \epsfysize=\columnwidth
\epsfbox{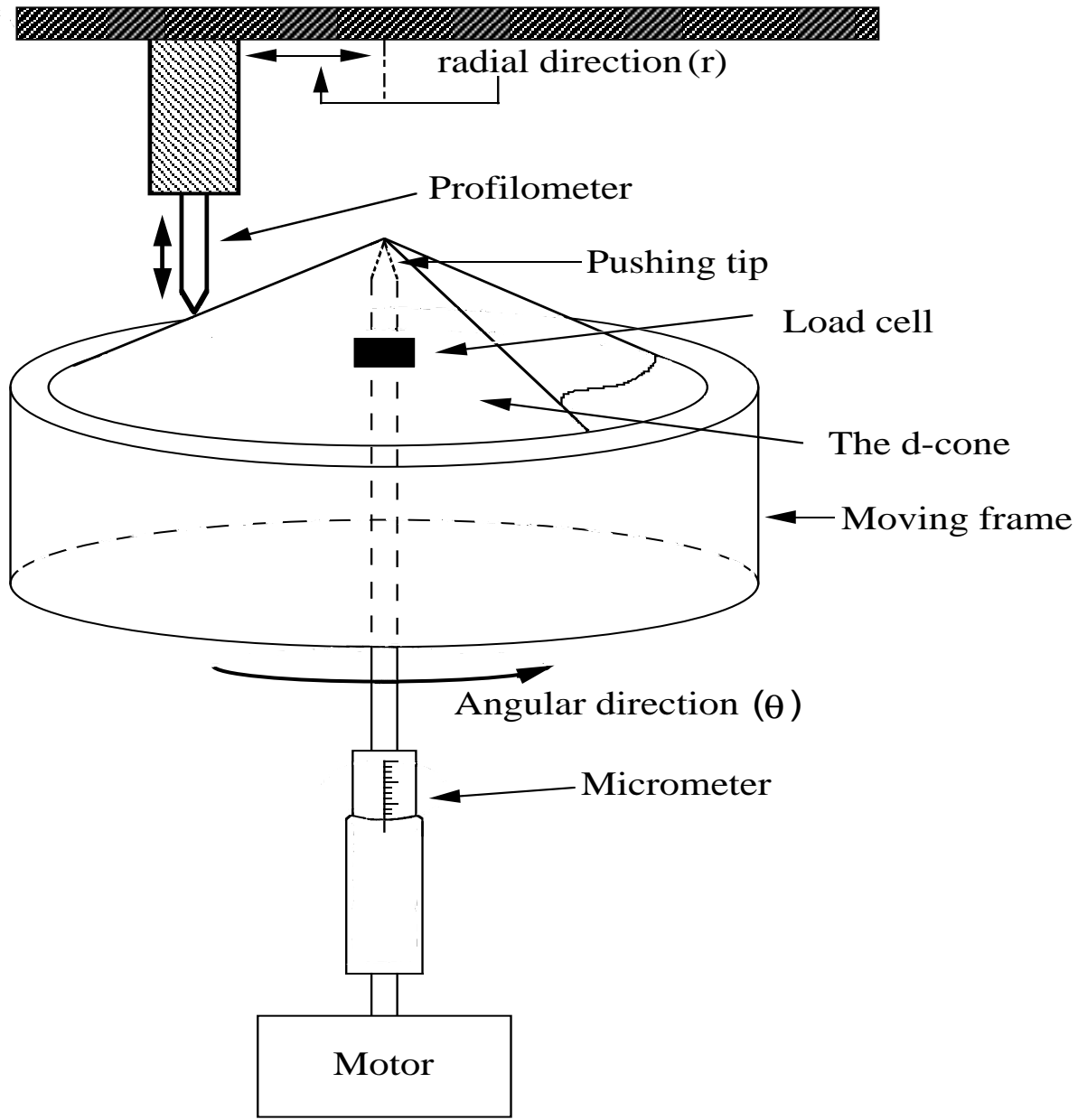}} 
\vskip .2in
\caption {The set up where the {\em d}-cone is performed and
where the profilometric measurements are achieved.}
 \label{disc}
\end{figure}

The {\em d}-cone is obtained on a thin circular plate by pushing a round tip
(0.5  mm diameter) centered at the plate principal axis. In this study we used 
circular plates made from both 0.05 and 0.1-mm-thick sheet 
(copper, brass, steel and transparencies); the results discussed are mainly from 
the 0.1-mm-thick sheets.  In order to allow the 
{\em d}-cone to form, when pushing the tip, we keep the sheet border  
free to move in a circular rigid  frame whose radius is 5 \% smaller than 
the sample radius $R_f $(Fig.~\ref{disc}). The opening angle 
$\phi $ of the {\em d}-cone (defined as the angle between the horizontal and 
the cone convex part generatrix) is varied by pushing the tip perpendicularly to the 
circular plate; the displacement $d$ is measured by a $10^{-2}$ mm precision 
micrometer. A miniature load cell is mounted under the pushing tip to allow
force measurements. 
The pushing tip is mounted on a rigid 20-cm-long and 1cm-thick steel pen-shaped 
cylinder. This bar is rigid enough to be inflexible when pushing the plate. 
A profilometric tip, mounted on the active part of a position sensor 
transducer, enables us to measure the sheet surface height with a precision 
of $10^{-2}$ mm. Two motors allow the tip to scan the whole {\em d}-cone 
surface; first, by moving the tip on a miniature automatic displacement guide 
mounted following the radial direction and second by rotating the frame around
its axis. Both the radial and angular directions are marked in Fig.~\ref{disc} as 
$(r)$ and $(\theta)$ respectively.  The measurements precision of the  
{\em d}-cone opening angle is approximately $7 \times 10^{-4} $ rad . 
The whole system is run by a PC computer equipped with A/D converter 
and data acquisition boards. In order to avoid stretching when a deformation $d$ is
imposed, a part of the plate  looses contact with the frame, giving rise to a 
concave region whose amplitude increases when increasing 
$d$ and whose location is randomly distributed on the plate.  If the pushing tip is 
deviated by a distance of the order of few
millimeters from the center, the characteristics of the sheet deflection are not changed, 
but its nucleation occurs in the closest region from the pushing 
tip to the frame border. In some cases, two or four {\em d}-cones appear one in front 
of the other. Pushing the 
plate further, one of the deflections amplitude increases while the others disappear.
The curvature at the ridge when two d-cones are present is half the curvature of the 
same d-cone if only one nucleates. When we approach the plastic transition
one of the d-cone desappears suddenly\cite{chaieb4}.

In the following we present local features of the buckled plate obtained by probing the
surface with a profilometer.
\protect
\section{Profiles}

\begin{figure}
\centerline{\epsfxsize=\columnwidth \epsfysize=\columnwidth 
\epsfbox{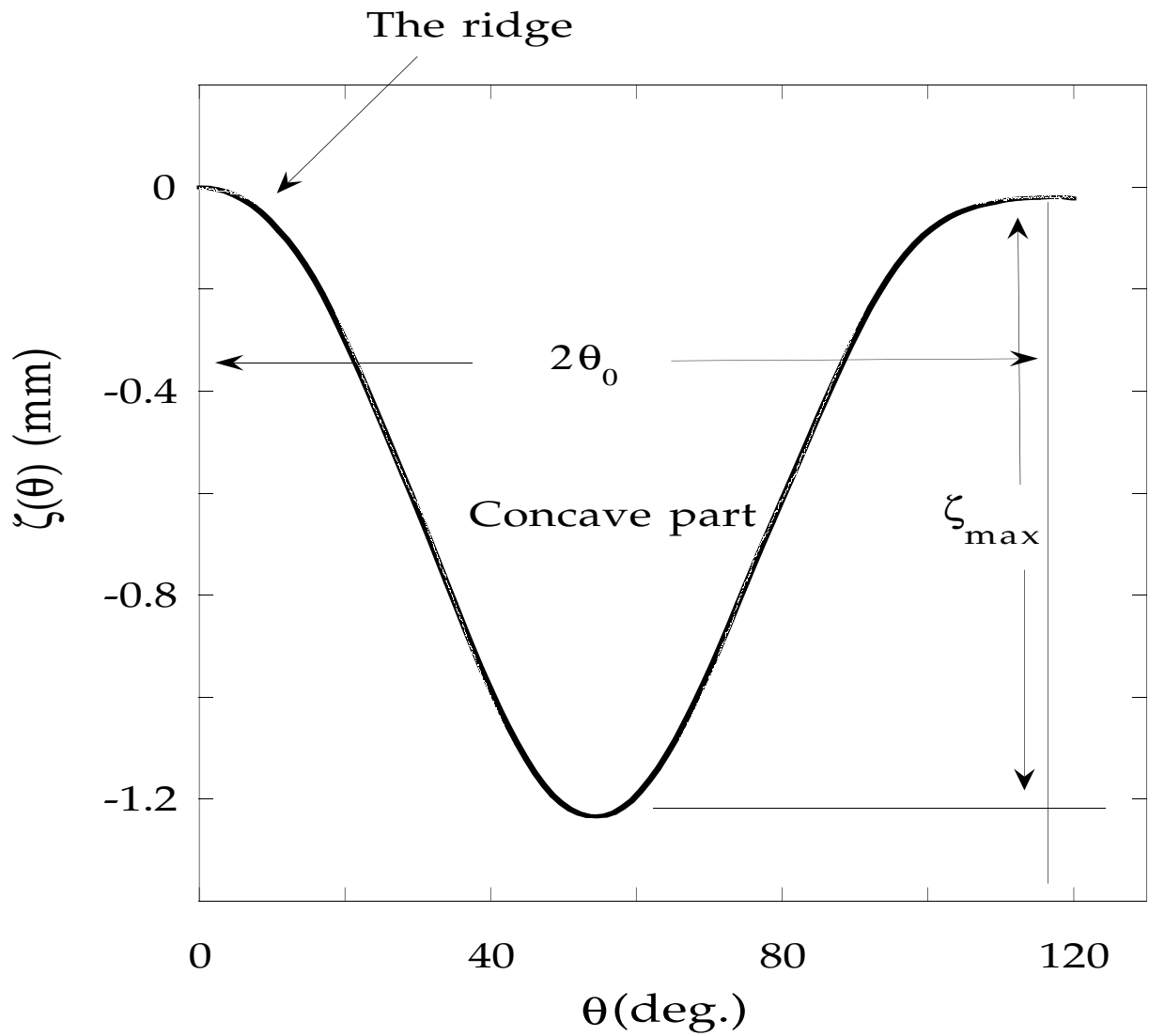}}
 \caption{The profile of the d-cone in cartesian coordinates. Notice the 
 angle $2\theta_0$ that measures the aperture angle between the points where 
 the plate looses contact with the frame}. 
\label{phitheta}
\end{figure}

\begin{figure}
\centerline{\epsfxsize=\columnwidth \epsfysize=\columnwidth 
\epsfbox{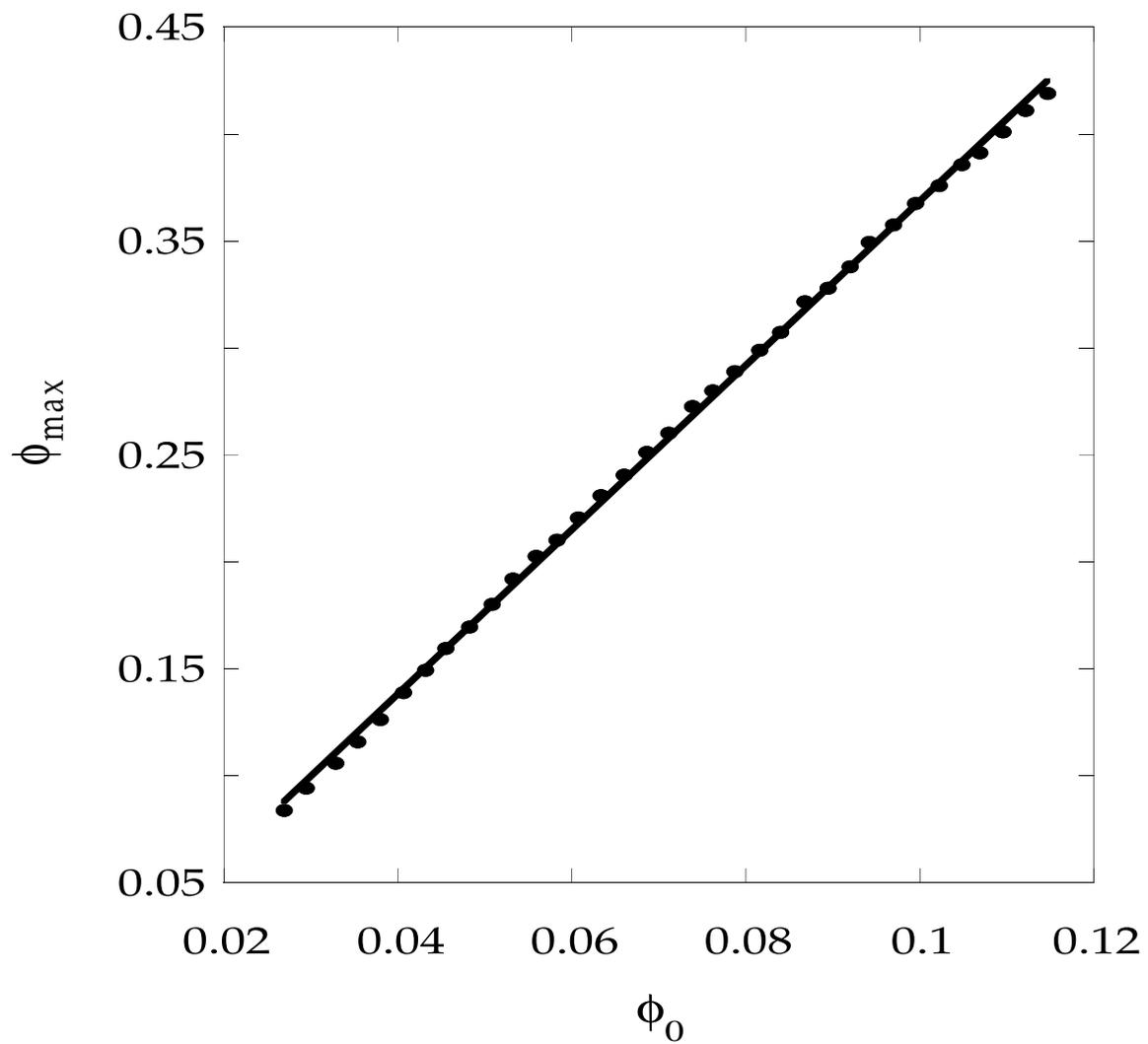}}
\caption{The plate maximum deflection as a function of the angle between the 
convex part and the horizontal. The distance to the pushing tip is 3mm.}
\label{deflec}
 \end{figure}

\begin{figure}
 \centerline{\epsfxsize=\columnwidth \epsfysize=\columnwidth \epsfbox{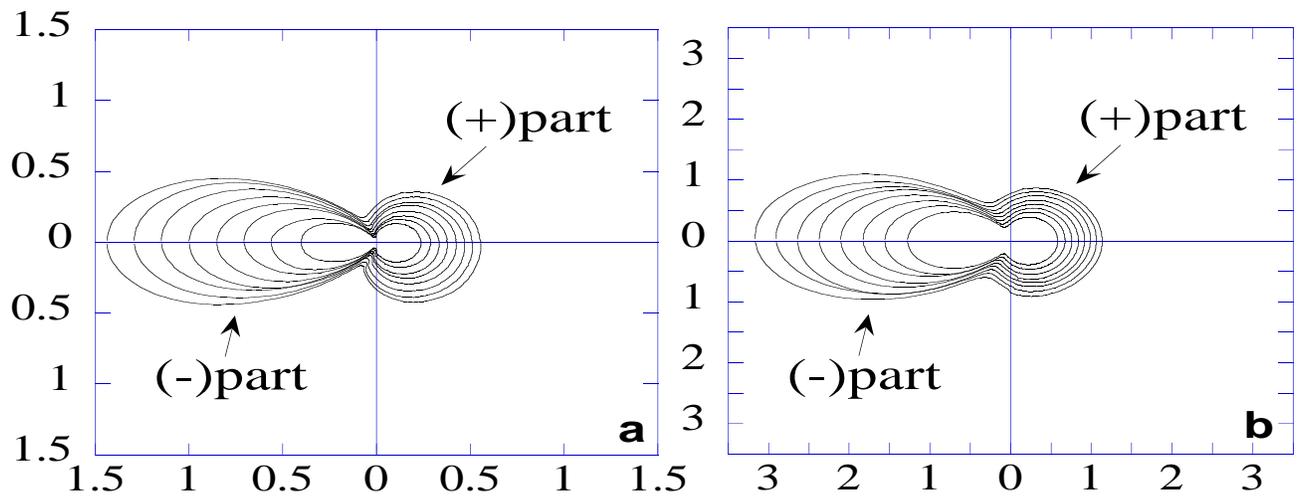}}
\caption{The profile of the sheet in polar coordinates vs. the angle $\theta$. a) d=1.41 mm;
 b) d=5.48 mm. The axis are in mm. The different curves correspondent to different 
 distances to the pushing tip.}
\label{prof}
 \end{figure}

In order to characterize the local geometry of the {\em d}-cone fully, 
we built a 
profilometer (shown in figure~\ref{disc}) which consists of a tip connected 
to a  transducer controlling the displacement of the tip.
The moving frame which supports the sample can have a 
very low angular velocity. The profilometer is able to resolve less than 
0.01 mm in the vertical ($\zeta$) and horizontal directions ($r,\theta$). 
Figure~\ref{phitheta}, displays the profile obtained at a given distance $r$ from the
singularity, and for a given deformation $d$.  From this profile, it is possible to 
measure the maximum deflection $\phi _{\text{max}}$ made by the concave 
region with the horizontal as a function of the angle $\phi_0$ made by the convex 
region and the horizontal. Also, from figure~\ref{phitheta} we measured the aperture 
angle of the deflection in the $\theta$ direction. This angle was found to be  
independent of the plate size and the deformation for a given geometry. Figure
~\ref{phitheta} displays the maximum deflection and how the angle $2\theta_0$ is 
measured. The almost horizontal line corresponds to the convex part of the cone.
The angle $2\theta_0$ is measured
when the plate is flattened and from the profiles where the cone is quenched in its 
configuration. 
It is important to notice that when the deformation is very high, the two procedures 
give two different angles. 
From this figure we measure the maximum deflection $\phi_{\text{max}}$, 
which is the lowest point in the figure~\ref{phitheta}. 
With ${\zeta_{\text{max}}(\theta)} = r \phi_{\text{max}}(\theta)$ where $r$ is the 
distance measured from the pushing tip. 
In figure~\ref{deflec}, we display  the dependence of $\phi_{\text{max}}$ versus the 
deformation expressed by the angle $\phi_0=d/R_f$. 
 The best linear fit to the  data in the figure above gives $\phi_{\text{max}} \sim 
 3.73\phi_0$. We will show in the following section that this selection of the aperture
 angle $2\theta_0$ and the maximum deflection as a function of the deformation
 can be explained by a minimization of the bending energy taking into account 
 that the deformation is isometric. 
 In polar coordinates, the plate profile looks like the ones displayed in figure~\ref{prof}. 
 Each one was obtained for a fixed distance from the pushing tip. Figure~\ref{prof}a, 
 corresponds to the {\em d}-cone profiles for a low tip displacement $d$ and for different
 distances $r$ from the pushing tip, and the 
 profiles for large deformations are shown in figure~\ref{prof}b. It is clear from this 
 figure that the convex part (circular curve in the figure) is off-centered. This shift of
 the d-cone tip will be explained as particular to d-cones made from a non zero thickness 
 sheets. 
 \protect
\section{Energy minimization of an isometric deformation}
In the following we  show 
that a simple model consisting of a minimization of the curvature energy taking 
into account that the deformation is isometric, can explain the above results and 
describe the d-cone far from the singularity. The general equation of a 
cone centered in O, in cylindrical coordinates, writes 
$z = r f(\theta)$. For convenience, we rewrite the parametric equation, 
$z=r\tan{\phi(\theta)}$ and $r=R\cos{\phi(\theta)}$ where $R$ is the distance 
to the tip and $\theta $ is the polar angle. A cone corresponds to a given 
function $\phi(\theta)$, where $\phi$ is defined as above. For a given 
deformation $\epsilon=d/R_f=\tan{\phi_0}$, where $d $ is the amount of the 
micrometer vertical displacement. If we write 
$\phi(\theta)=\phi_0$ for $|\theta| > \theta_0$ and,
\begin{equation}
\phi(\theta)=\phi_0 + \gamma\bigg(1+\cos{\pi{\theta \over \theta_0}}\bigg)
\text{ for }|\theta| < \theta_0.
\label{phi}
\end{equation}
The function $\phi(\theta)$ defines then a cone that remains in contact with 
the circular frame for $|\theta| > \theta_0$.  The {\em d}-cone is detached 
from the plate over an angle equal to $2\theta_0$ which corresponds to the 
deflection. We assume that $d$ is small and that $\gamma$ and $\phi_0$ are 
of the same order of magnitude. To the first order, the total curvature of the 
surface then reduces to $\kappa = (\phi + \phi'')/R$. The corresponding 
energy $E_{\kappa}$ (per unit of R) is :

\begin{eqnarray}
E_{\kappa}&= &{K \over 2} \int_{-\pi}^{\pi} 
{(\phi + \phi'')^2 \over R^2}R\sqrt{\cos^2{\phi}+\phi'^2}\,d\theta \nonumber \\
&\sim &{K \over 2 R}\Biggl[2\pi\phi_0^2 
+4\theta_0\phi_0\gamma
+\bigg(
3\theta_0-{2\pi^2 \over \theta_0}+{\pi^4 \over \theta_0^3}
\bigg)
\gamma^2\Biggr]
\label{energy}
\end{eqnarray}

For an unstretchable plate, the length $L$ of the corresponding line at 
$R=constant$ must be equal to $2\pi R$ so that :

\begin{eqnarray}
2 \pi R &= &
L =
\int_{-\pi}^{\pi} R\sqrt{\cos^2{\phi}+\phi'^2}\,d\theta \nonumber \\
&\sim&
\Biggl[2\pi-\pi \phi_0^2
-2\theta_0\phi_0\gamma
-\bigg({3 \theta_0 \over 2}-{\pi^2 \over 2 \theta_0 } \bigg)\gamma^2\Biggr]R 
\label{length}
\end{eqnarray}

Equation~\ref{length} gives $\gamma$ as a function of $\phi_0$
and $\theta_0$. Replacing $\gamma$ by its value in equation~\ref{energy} 
and minimizing $E_{\kappa}$ with respect to
$\theta_0$, one finds $2\theta_0 \sim 2.09$~rad $ \sim120^{\circ}$ and
$\gamma \sim 1.38~\phi_0$.  This last relation confirms our assumption that 
$\gamma$ and $\phi_0$ are of the same order of magnitude. An exact solution 
of a similar problem gives $2\theta_0 \sim 140^{\circ}$ \cite{maha}. As pointed out above
our result is valid for low deformations, when the aperture angle is the same when
measured in the plate frame and the laboratory frame.
The theoretical value of the aperture angle $2\theta_0$ is in good agreement 
with the experiment (Fig~\ref{phitheta}). The aperture angle between the points 
where the plate looses contact with the frame is about $(110\pm 5)^{\circ}$. 
Experimentally, and even for large $d$, the aperture angle depends very slowly 
on $d$ when measured in the plate coordinates. The theoretical maximal deflection 
angle $\phi_{\text{max}}=\phi(0)$ is proportional to $\phi_0$ and equals 
$\phi(0)=(\phi_0+2\gamma)=3.76~\phi_0$. This result is also in good agreement 
with the experimental data even at large $d$, since the best fit in Fig.~\ref{deflec} 
gives $\phi(0) =3.73~\phi_0$. It is worth noting that these results are valid within
the range of deformation so that if released the cone recover its shape. However the 
global shape of the d-cone, as shown in figures \ref{prof} and \ref{phitheta}, is purely 
geometric as we will show later.
\protect
\section{Local properties of the developable cone}
In this section we present the general features of the surface of a developable cone. From
the profiles presented in the previous section, we can retrieve the local curvature of the
concave part and the ridge (the region separating the concave region and the convex 
region) as well as the curvature of the crescent shape at the pushing tip. Also we will show
that  developable cone made from a thick sheet ($h \ne 0$) is not centered at the
pushing tip. We will then  define a function called the d-cone anisotropy which measures 
how far is the ``experimental'' d-cone from the theoretical d-cone in geometrical terms.
\protect
\subsection{Anisotropy}
If one looks carefully to the profiles in figure~\ref{prof}, one notices that for small 
deformations the profiles are sharper than the ones corresponding to large 
deformations. Also the origin of the circular part of the profiles is not centered
at the coordinates origin but shifted to the right of figure~\ref{prof}. This 
shift is due to an anisotropy of the plate. This anisotropy is due to the fact that when 
pushing the plate to make the deflection that corresponds to the d-cone, the tip of the
d-cone obtained is then shifted to allow the deflection to form. It costs more
energy, at least by making a curvature, to produce a small point with a divergent 
curvature than a lrage deflection with small curvature. Hence, to save energy 
necessary to make a sharp vertex,the singularity is rejected out of the plate by a 
distance that is, for low deformations, equal to the frame radius, so that the plate 
looks smooth.The generatrices no longer meet at the pushing tip. 
This shift can be described with a simple geometrical model. It can be quantified 
by measuring what we called the {\em d}-cone {\em anisotropy} $\mathcal{A}$ defined 
as the ratio $(\zeta (\pi) -\zeta (\pi /2))/\zeta (\pi)$, where $\zeta (\theta) $ is the height 
of the sheet  measured at the polar angle $\theta $ and $\pi$ corresponds to the
very right point in the profiles of figure~\ref {prof}. In the following section we present 
how by cutting an ``effectif'' cone with a plane we can find the distance we called $x_s$ by
which the tip has moved.
\protect
\subsection{Geometrical model}
\begin{figure}
\centerline{\epsfxsize=\columnwidth \epsfysize=\columnwidth \epsfbox{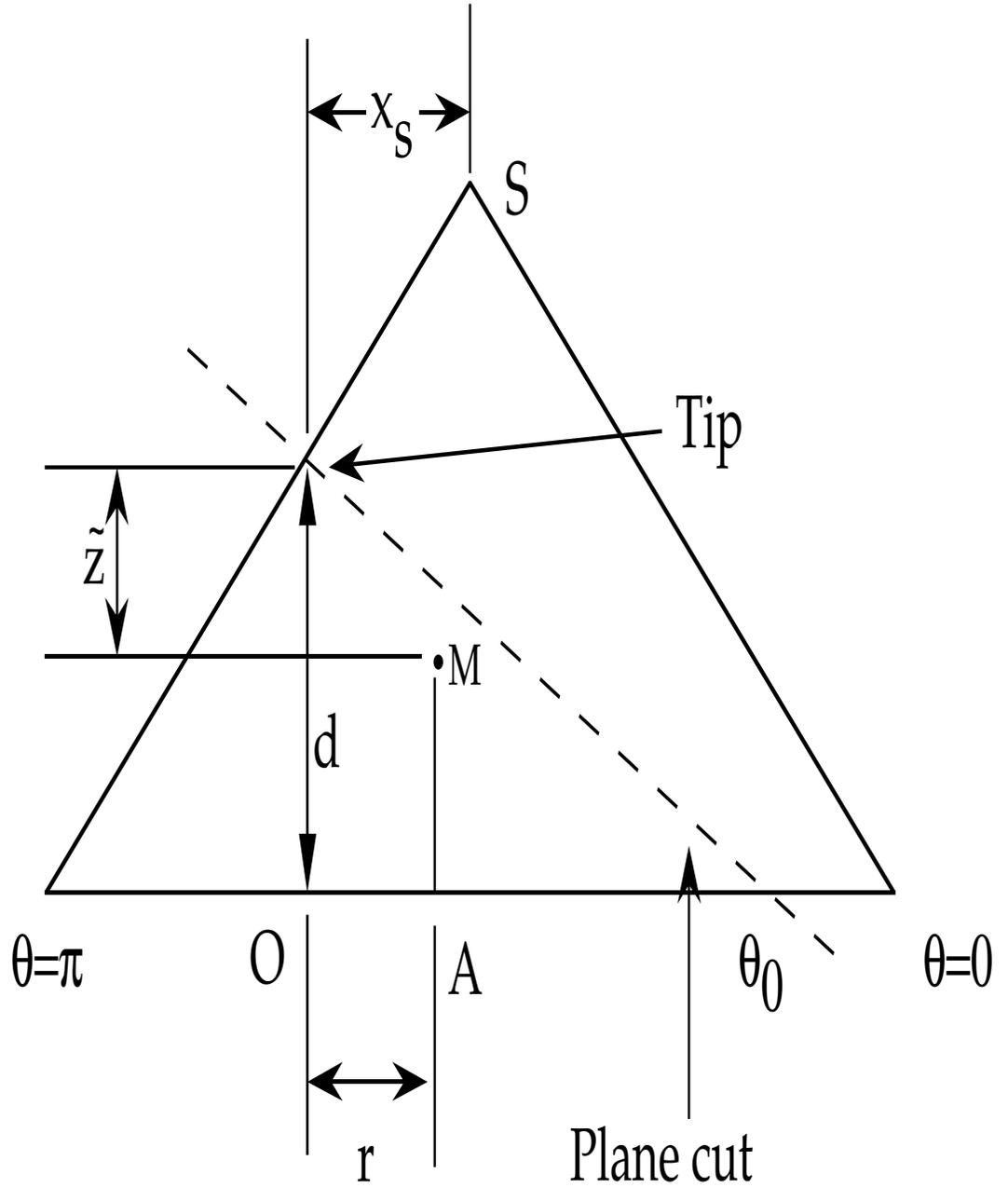}}
\caption{Geometrical construction of the d-cone obtained after a cut by a plane of a 
perfect cone}
\label{aniso}
\end{figure}

If we look at the d-cone ridges, where the crescent appear, we find that they define 
a plane. Furthermore, the generatrices 
do not meet at the d-cone pushing tip. From figure~\ref{prof}, we notice also that the
circular part, that is the convex region, is not centered at the coordinates origin.
The obtained cone is as if it is cut by a plane defining then an aperture angle $2\theta_0$. 
In figure~\ref{aniso}, we show the geometrical location where the plane and the cone 
meet. In the following we show the origin of this ``anisotropy''.

If S, A and M belong to the cone,they are related by 
\begin{equation}
\roarrow{SM}=\lambda \roarrow{SA}
\label{homotetie}
\end{equation}
Where S is the tip and A is on the line defined by the intersection of the moving 
frame and the plate of fig.~\ref{disc}. If O is on the pushing line then, 
$\roarrow{OA}=R(\cos \theta \,\vec{i} + \sin \theta \,\vec{j})$.
Also we have: 
\begin{eqnarray}
\roarrow{SA}&=& (R\cos \theta-x_s)\vec{i}+R\sin \theta \vec{j} -z_s \vec{k} \nonumber \\
\roarrow{SM}&=&(x_m-x_s)\vec{i}+(y_m-y_s)\vec{j}+(z_m-z_s)\vec{k}.
\end{eqnarray}
From equation~\ref{homotetie} we have:
\begin{eqnarray}
x_m & = &  x_s+\lambda (R\cos \theta-x_s)    \nonumber\\
y_m  & = & \lambda R \sin \theta  \nonumber \\
z_m & = & (1-\lambda)z_s
\end{eqnarray} 

In the previous relations,  $(x_m, y_m,z_m)$ are the coordinates of M, and 
$(x_s,y_s,z_s)$ are the coordinates of S. In our case R is the frame radius.
One needs to find a relation between $\lambda$ and $\theta$, in fact this can be easily 
achieved by calculating a distance $r^2 = x_m^2+y_m^2$ on the cone.
At $r$ constant we have:
\begin{eqnarray}
& &{\lambda}^2 ((R\cos \theta-x_s)^2+R^2{\sin}^2 \theta)  \nonumber\\ 
&+& 2\lambda x_s(R\cos \theta-x_s)+x_s^2-r^2=0
\label{lambda}
\end{eqnarray}
This equation gives us $\lambda (\theta)$ for a given $r$.
The height is now given by $z_m=(1-\lambda (\theta))z_s$. It is more convenient to
reverse the $z$ axis and consider a direct cone so that the generatrices are in 
the half-plane $y>0 $ axis. We define then $z_m=d-{\tilde{z}_m}$ (The profiles 
in figure~\ref{prof} are obtained in reversed axis) and obtain:
\begin{equation}
{\tilde{z}_m}=\lambda (\theta) d+(1-\lambda (\theta)){\tilde{z}_s}
\end{equation}

\begin{figure}
\centerline{\epsfxsize=\columnwidth \epsfysize=\columnwidth  
\epsfbox{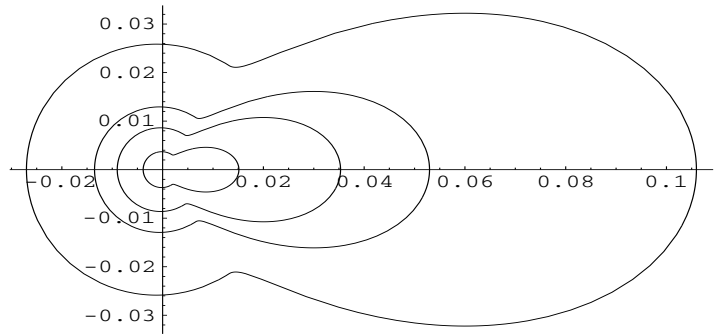}}
\caption{The d-cone profiles obtained from the geometrical model. Each curve 
correspond to a distance from the origin. The distance $x_s$ is set to 1. The frame radius 
is set to 37 mm and the thickness to 0.1 mm. The different distances are $r$=2,4,6,12 
(scaled units). }
\label{profmat}
\end{figure}

\begin{figure}
\centerline{\epsfxsize=\columnwidth \epsfysize=\columnwidth  
\epsfbox{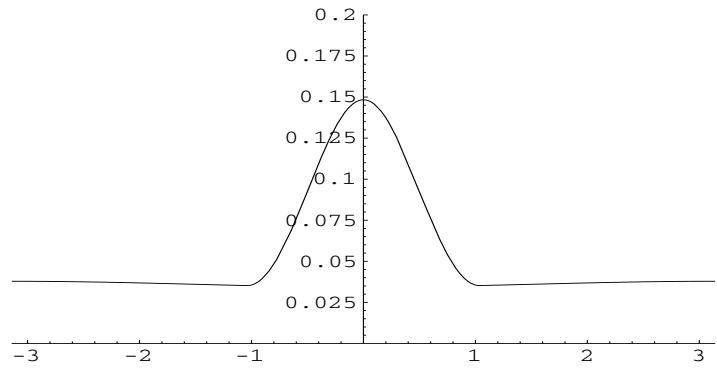}}
\caption{The plate height $\tilde{z}_m$ obtained from the geometrical model. The parameters 
are the same as in figure \ref{profmat}. The distance to the tip is 2 scaled units. }
\label{phithetamat}
\end{figure}

If we plot the lines $\tilde{z}(\theta)$ we recover the experimental profiles in 
figure~\ref{prof}. In figure \ref{profmat}, we display the profiles calculated from
this model.  The different profiles correspond to different distances $r$ from the
origin. The deformation is measured by calculating $x_s$. The frame radius is
the one used in the experiments and the thickness is set to 0.1 mm. From this 
figure we notice that the shape of the d-cone can be obtained from a simple
geometrical model with the ansatz \ref{phi}.
In figure \ref{phithetamat}, we plot the height $\tilde{z}_m$ versus the polar angle
$\theta$. From this figure the opening angle is equal to 114 degres. The aperture angle
is selected geometrically.

The cut in figure~\ref{aniso} defines then a hyperbola whose equation is found as 
the following:
The plane when cutting the cone defines an angle $\alpha$, so that 
$\tan \alpha =x_m/(z_m-d)=-(R \cos \theta_0 )/d$. The intersection of the plane and 
the cone is given by:
\begin{eqnarray}
X &=&\lambda R \sin \theta  \nonumber \\
Y&=&-(1/\sin \alpha)\Bigl(x_s (1-\lambda)+\lambda R \cos \theta \Bigr)
\label{plane}
\end{eqnarray}

with 
\begin{equation}
\lambda (t) = ((z_s -d) \tan \alpha -x_s )/(R \cos t -x_s + z_s \tan \alpha)
\label{lam}
\end{equation}

Eliminating $\lambda (t)$ between equations \ref{plane} and \ref{lam} we find
\begin{eqnarray}
& &X^2 \Bigl(1+\frac{x_s}{R}\Bigr)^2 \Bigl(d^2 +R^2 \cos ^2 \theta_0\Bigr) \nonumber \\
&+&Y^2 \Bigl( (x_s+(R+x_s)\cos \theta_0)^2 -R^2\Bigr) \nonumber \\
&-&2Yx_s (R+x_s)(1+\cos \theta_0)-\sqrt{d^2+R^2 \cos^2 \theta_0} =0
\label{hyperbo}
\end{eqnarray}

Equation \ref{hyperbo} is the equation of a hyperbola whose curvature at the tip is given 
by:
\begin{equation}
 \kappa = \frac{(R+x_s)\sqrt{d^2 + R^2 \cos^2 \theta_0}}{R^2 x_s (1+\cos \theta_0)}
{\sim} \frac{1}{2x_s} \: (x_s << R)
\end{equation}
 We showed that with this one dimensional geometrical model, we can characterize the size 
 of the singularity and found that it belongs to a hyperbola defined as the 
 intersection of a plane with a perfect cone. Physically, the size of the singularity 
 is due to the fact that for a ``real'' sheet, it is 
 energetically favorable to create a deflection by bending and rejecting the singular point 
 far away from the tip and the {\em d}-cone obtained does not have 
 a singular tip or a vertex. It is beyond this model to explain how the crescent form
 and how its curvature depends on the deformation, here the parameter $x_s$ is equivalent 
 to the experimental deformation.
 For small deformations the 
 deflection exists but the plate is smooth everywhere as if the size of the singularity is all
 over the surface.
\protect
\subsection{The shift from the anisotropy}
\begin{figure}
\centerline{\epsfxsize=\columnwidth \epsfysize=\columnwidth  \epsfbox{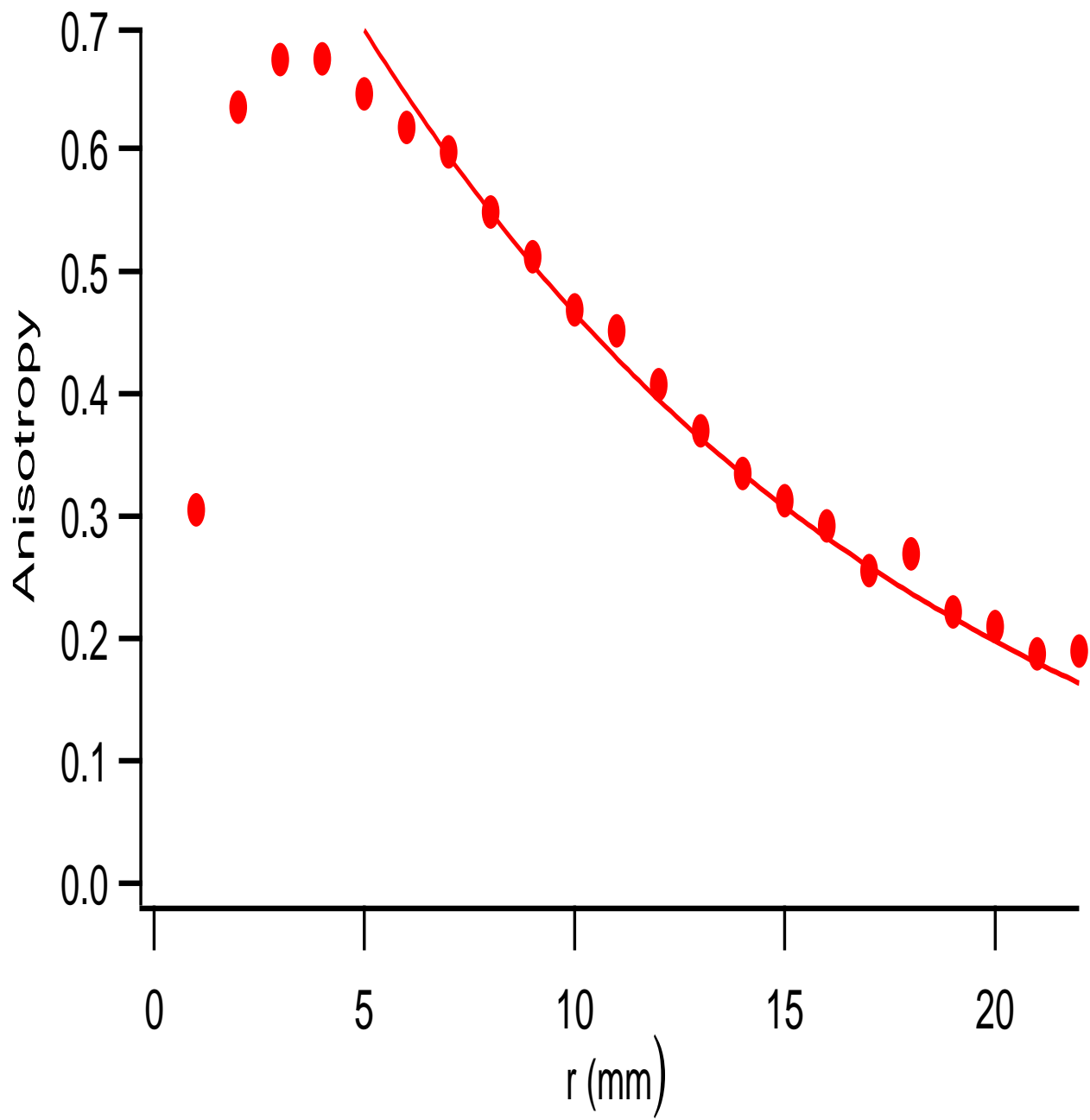}}
\caption{The anisotropy $\mathcal{A}$ as a function of the distance r. The deformation is
1.5 mm .}
\label{ani_r}
\end{figure}

From figure~\ref{aniso}, we can define an angle $\alpha$ given by $\tan \alpha=d/R$ 
so that $z_s=-x_s \tan \alpha$. We define the anisotropy $ \mathcal{A}$$(r,R_f,x_s)$
keeping in mind that the deflection is centered at  $\theta =0$.
\begin{eqnarray}
{\em A}&=& (\tilde{z}(\pi) -\tilde{z} (\pi /2))/\tilde{z}  (\pi)  \nonumber \\
&=&\Bigl(-rR +x_s (r-R)+\sqrt{B}\Bigr)/r(x_s - R)
\label{AA}
\end{eqnarray}
Where $B=(r^2R^2-x_s^2 (r^2-R^2))$.
The distance $x_s$ is obtained by measuring the heights
from the profiles like the ones depicted in figure~\ref{prof}, and fitting the data with
expression \ref{AA} giving the anisotropy versus the frame radius $R_f$, the distance 
$r$ and the distance $x_s$ which is deduced from the fit.
In figure~\ref{ani_r}, we show an example of the anisotropy measured from the profiles, 
and the line is the best fit with the formula \ref{AA} for a given deformation. It is
note worthy that the anisotropy, found experimentaly, decreases when we go away from 
the sigularity. This effect is due to the fact that close to the singularity, the plate 
suffer a stress focusing and an irreversible deformation would take place if we increase
the deformation. Further away from the singularity, the landscape is smoother and the 
plate is no longer anisotropic.

In the next section we will show that, the local curvature of the 
concave part does not follow a law of the form 1/r, but the coordinates are
shifted by a value we call $r_s$ which depends on the deformation. 
The shift in coordinates origin $r_s$ is correlated to
the displacement of the singularity $x_s$. We will show that due to stress focusing, this 
distance decrease when the deformation is increased .
\protect
\subsection{Size of the singularity and stress focusing}
\begin{figure}
\centerline{\epsfxsize=\columnwidth \epsfysize=\columnwidth 
\epsfbox{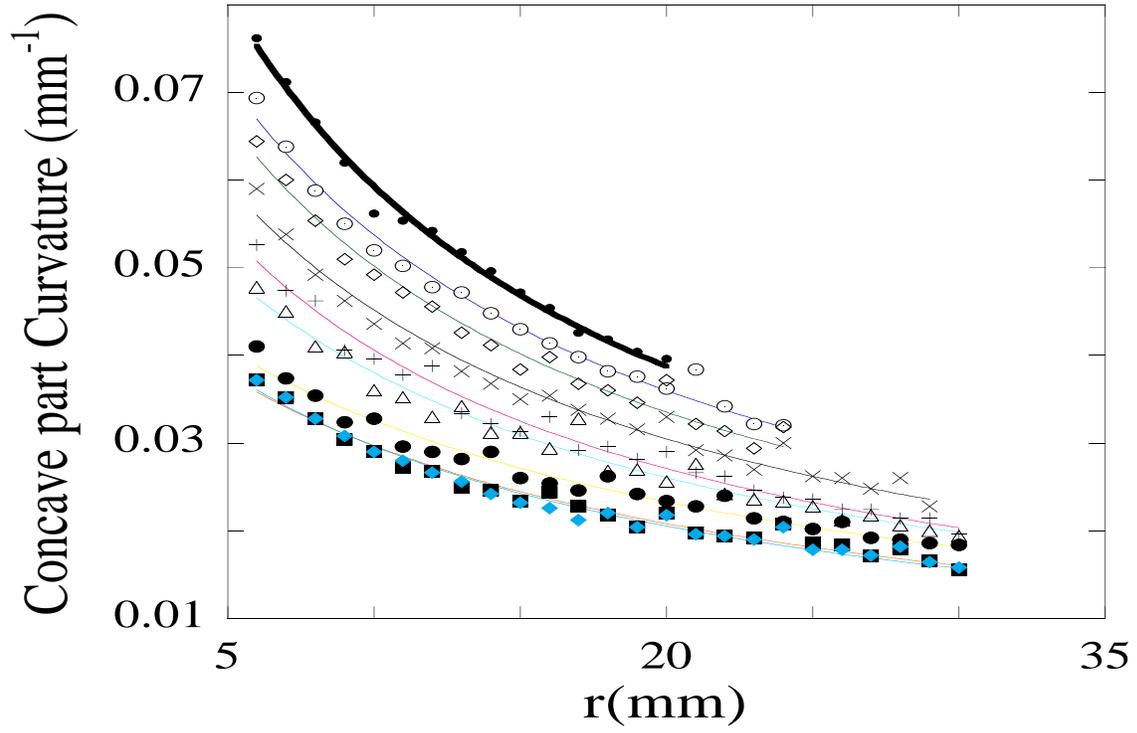}}
\caption{The concave-part local curvature vs. the distance to the singularity, and for 
different small $d$. The line is a best fit to the function $1/(r+r_s)$.}
\label{curv}
\end{figure}

\begin{figure}
\centerline{\epsfxsize=\columnwidth \epsfysize=\columnwidth 
\epsfbox{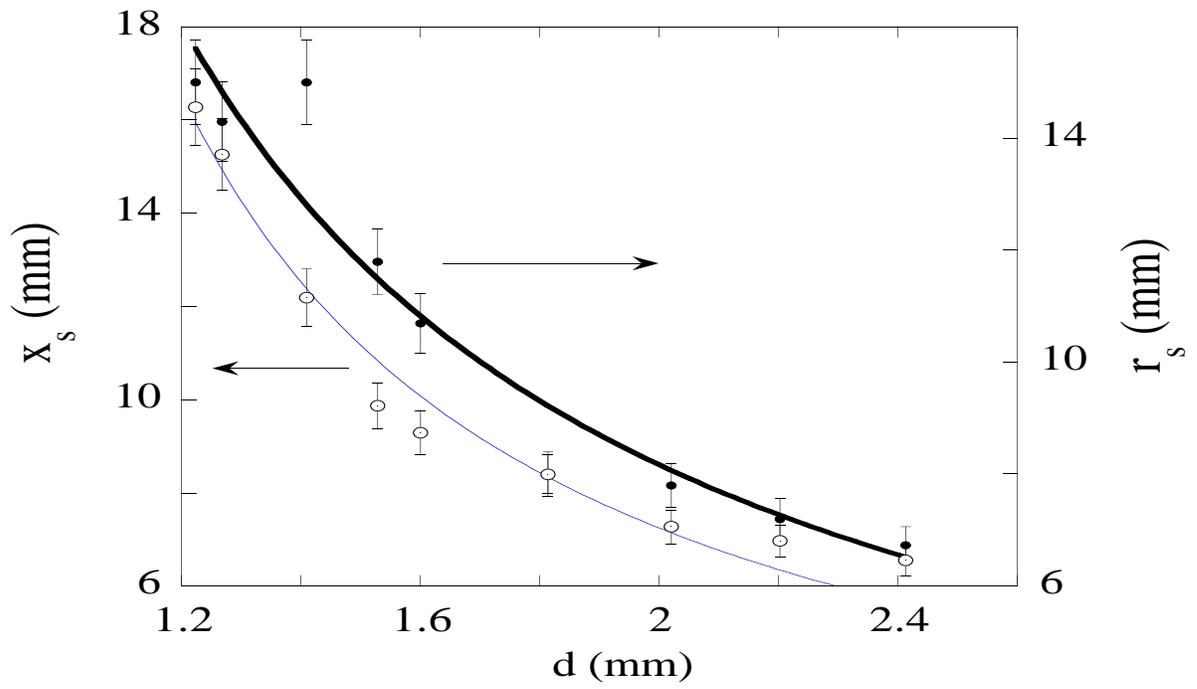}}
\caption{The shift distance $r_s$ and the singularity displacement $x_s$ vs. d}
\label{xsrs}
\end{figure}

We have measured  the local curvature of the concave region, and found that 
for small deformation, the radius of curvature is linear with the distance from 
the tip, but the origin is shifted by $r_s$.  It is well known that at each point of a 
perfect cone, there is no curvature towards the vertex. The curvature decreases like 
$1/r$, where $r$ is the distance from the vertex. In figure~\ref{curv}, we display the 
local curvature versus the distance from the pushing tip. The line
is the best fit to a function of the form $1/(r+r_s)$. 
From the figure~\ref{curv}, the origine of coordinates is not centered at the
pushig tip, but it is shifted by a distance $r_s$. This distance is found to be a decreasing
function of the deformation as well as the distance $x_s$. It is tempting to think that
the shift in the coordinates origine $r_s$ is another way the d-cone avoid making
a singular point and moved it  out of the plane. Notice that $r_s$ however, is not exactly 
equal to $x_s$.
As a result of the stress focusing, the size of the singularity which is of 
the size of the frame radius at small deformations, decreases and also the 
shift in the coordinates collapse to the same origin as the cone whose tip is at 
S (fig.~\ref{aniso}). In figure~\ref{xsrs}, we depict $x_s$ and $r_s$ versus the 
deformation. 

Notice that both $r_s$ and $x_s$ decrease when the 
deformation is increased. As the bending rigidity goes like $h^3$ where $h$ is small,
the creation of a punctual
singularity cost more energy than a simple deflection by bending the surface,
from figure~\ref{xsrs} we notice that at small deformations the singularity is rejected
to infinity and the whole surface is bent and the size of the singularity is of the order of
magnitude of $R_f$ or even larger. When the deformation increases the size of the 
singularity decreases by decreasing the distances $x_s$ and $r_s$. 
The stress focusing can be seen as a decrease in the singularity size by strain localization.
\protect
\subsection{Curvature at high deformation: Stretching effect}
\begin{figure}
\centerline{\epsfxsize=\columnwidth \epsfysize=\columnwidth \epsfbox{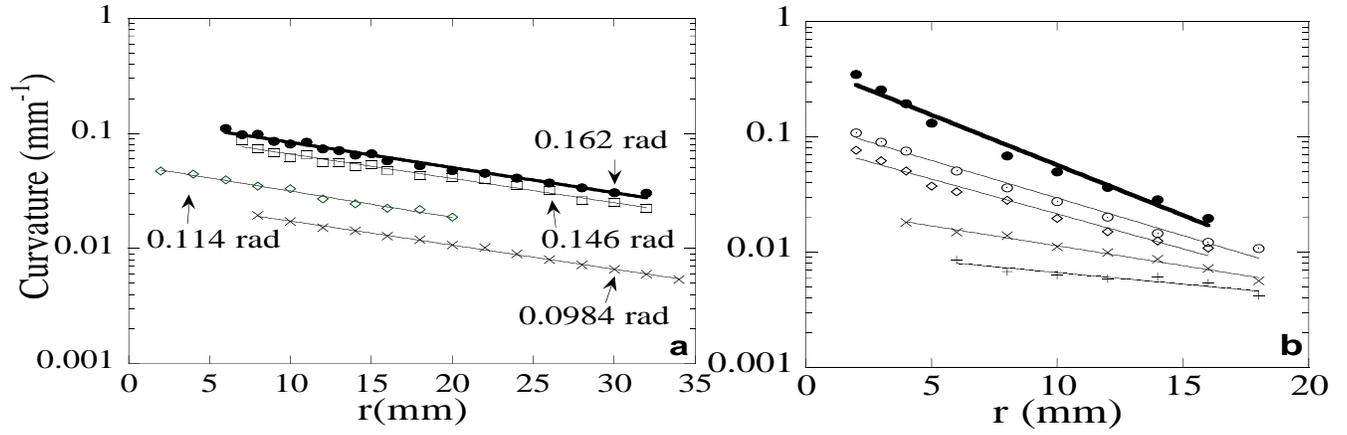}}
\caption{The d-cone local curvature vs. the distance r from the pushing tip for
different deformations. (a) the local curvature of the concave part, where $r_c$ is constant 
and equal to 20 mm. (b) the ridge local curvature, where $r_c$ decreases when $d$ 
increases. In the figure we reported the values of the opening angle and not d}
\label{curv-str}
\end{figure}

As we increase the deformation a line with a different texture from the rest of the plate
appear at the ridges, and the curvature increases. In order to characterize this transition
we measured the curvature at high deformation both in the concave region 
and at the ridge. 
In figure~\ref{curv-str} we depict the curvature of the concave region (a) and 
of the ridge line (b). Each line corresponds to a given deformation.

From figure~\ref{curv-str} we notice that the curvature is no longer of the form $1/r$
but it decreases exponentially with the distance like $C_0 e^{(-r/r_c)}$  where $r_c$
is a characteristic  distance. In figure~\ref{curv-str}(a) the 
characteristic  distance $r_c$ is constant versus the deformation whereas on the ridge
fig.~\ref{curv-str} $r_c$ decreases with increasing the deformation. This behaviour is
due to the fact that the crescent due to the scar appears only on the ridge. In other words,
the plastic deformation is felt on the ridge where the plate is folded and where the stress is 
concentrated. The slope of the top line in figure~\ref{curv-str}a, reaches a 
value that corresponds to the radius at which the yield limit of a 0.1-mm-thick copper 
sheet is exceeded and where a permanent scar appears \cite{chaieb1}. 

When one folds a sheet of paper to make a developable cone, one notices that the 
curvarture is not exponential but decreases algebraically. In this case
the deviation of the curvature from 1/r 
behavior to an exponential is due to the fact that at large $d$, the yield limit of 
the material is exceeded and stretching starts to be more important than pure 
bending because, contrary to a free sheet, this one is squeezed within a circular frame. 
This is why the stretching effects are noticeable and  the near the borders the deflection 
looks rather more flatened than if the d-cone is borders-free.
As a first approximation we assume that concave part is an isolated 
stripe. By further pushing the plate beyond the yield limit, the stripe starts to 
bend and the region near the singularity, at the pushing tip, suffers stretching. 
Following\cite{landau}, if we include stretching in the energy balance we find 
that the stripe local curvature decreases like an exponential, and the cutoff 
distance decreases by increasing the height of the sheet, that is by pushing 
the plate \cite{lobko1}. The curve giving the curvature versus the distance 
for a {\em d}-cone made of a 0.05-mm-thick sheet, gives a cutoff distance that 
is the half of the one for a 0.1-mm-thick sheet. We noticed no qualitative changes 
between the two plates, and we believe that the cutoff distance is a linear function 
versus the plate thickness $h$. 

Another way to characterize the singularity size, 
is to measure the properties of the crescent shape observed at the pushing tip.
\protect
\subsection{The crescent singularity}
Crumpling a thin sheet or better a transparency, leaves a scars that looks like crescents. 
These crescents are the result of stress focusing. One wonders why the stress when focused
does not leave pointlike scars. This is due, as discussed above, to the fact that making 
a singularity whose radius of curvature is of the order of $h$ costs more energy than 
pure bending. Instead, it is preferable
to make a crescent whose spatial extension is orders of magnitude larger than $h$. It is
then of a great importance to measure the size of the crescent 
 left after crumpling. In our experiment, we measured the radius of curvature $R_c$ of the 
crescent as a function of the deformation for small and large deformations.
\protect
\subsubsection{Radius of the parabola for low deformations}
\begin{figure}
\centerline{\epsfxsize=\columnwidth \epsfysize =\columnwidth \epsfbox{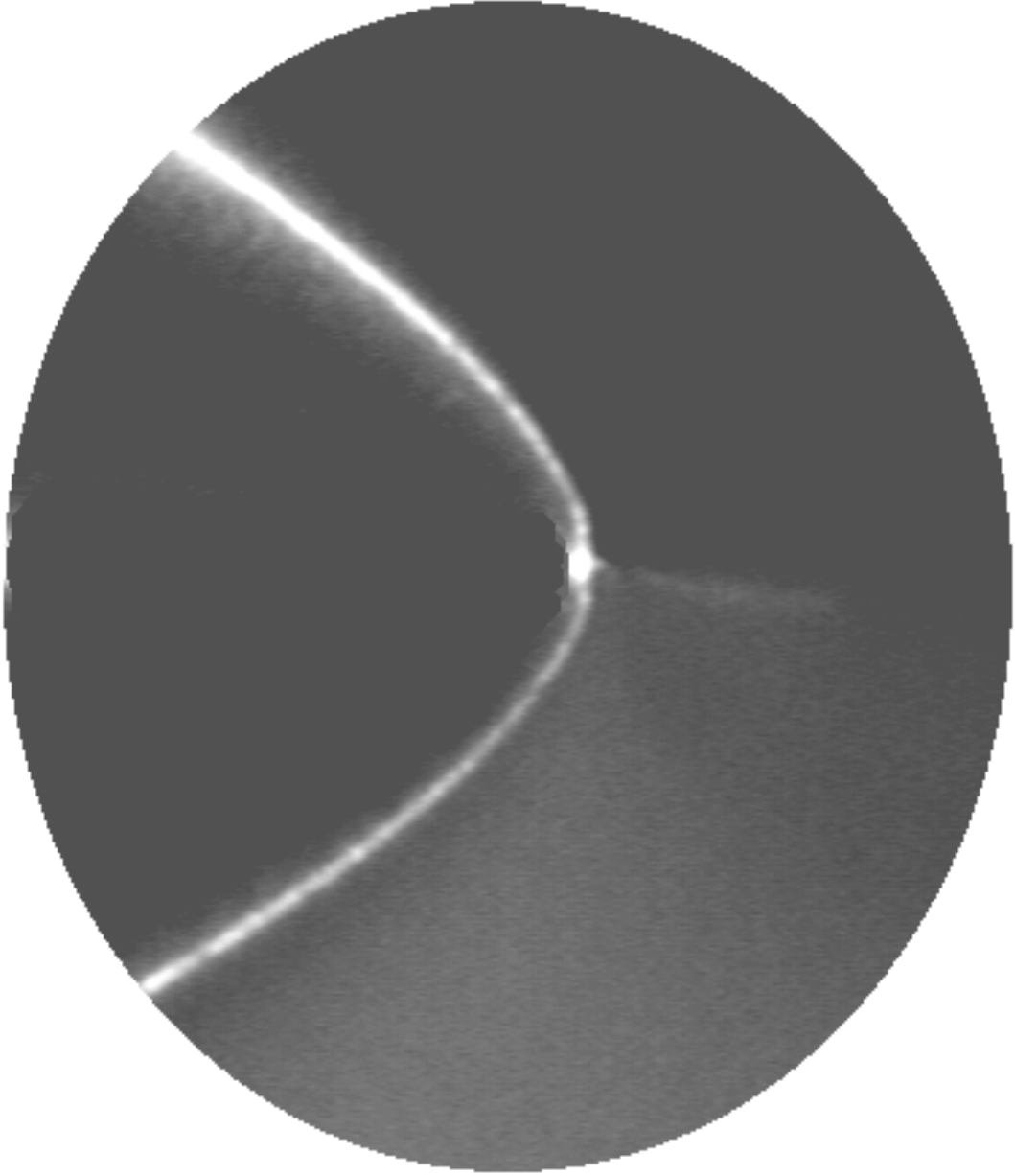}}
\caption{The d-cone observed from above, and illuminated perpendicularly to the
plane of the bright parabola. The deformation
is about 0.3 mm. $R_f$=22.5 mm}
\label{dc-faib}
\end{figure}

 To measure the crescent radius of curvature, we illuminate 
the d-cone from above so that the light beam is perpendicular to the ridge. The ridge
reflect more light than the rest of the plate as its texture is changed.
Figure ~\ref{dc-faib} displays the d-cone for a small deformation. Notice the parabolic 
shape of the bright line separating the convex region from the concave region, and define 
an angle smaller than $2 \theta_0$. The image looks oval, because as the plane defined by 
the parabola makes an angle $\pi /2-\alpha = \arctan \epsilon$  with the horizontal, 
the frame is twisted by the same angle so that the light beam is perpendicular to the 
parabola ($\alpha $ is defined in figure~\ref{aniso} where the opening angle is
exaggerated).

We digitalize the image and collect the points belonging to the bright line. With is numerical
procedure the dark part
of the figure is not counted. We then have a parabola whose radius of curvature can be 
easily calculated by fitting the obtained curve to a second order polynomial 
\cite{chaieb1,chaieb2}.

\begin{figure}
\centerline{\epsfxsize=\columnwidth \epsfysize=\columnwidth 
\epsfbox{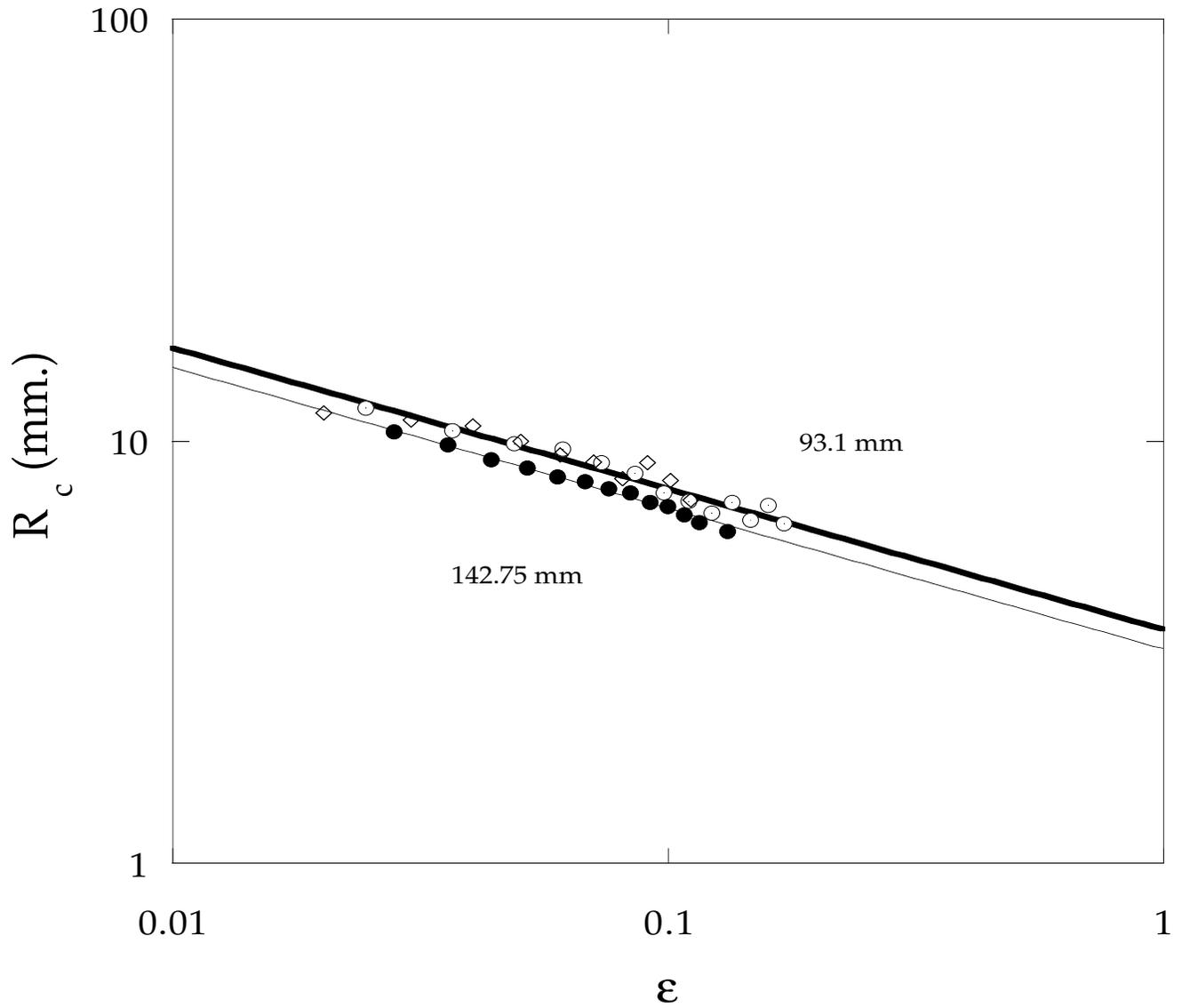}}
\caption{The radius of curvature of the crescent for small $\epsilon$. The line is
a best fit to the power law $\epsilon ^{-1/3}$}
\label{r-faib}
\end{figure}

Figure~\ref{r-faib} depicts the radius of curvature of the bright parabola in 
figure~\ref{dc-faib} as a function of $\epsilon = d/R_f$, or the angle between the convex 
region and the horizontal. We show the data in a linear scale for the sake of clarity. 
We observe that the radius of cuvature of the crescent scales like ${\epsilon}^{-1/3}$, where 
$\epsilon $ is defined above. In this regime, the deflection is just moving in 
the vertical deflection as we deform the plate. This is due to the reaction force experienced
by the plate at the point where the plate looses contact with the frame.

\protect
\subsubsection{Radius of the crescent at high deformations}
\begin{figure}
\centerline{\epsfxsize=\columnwidth \epsfysize=\columnwidth \epsfbox{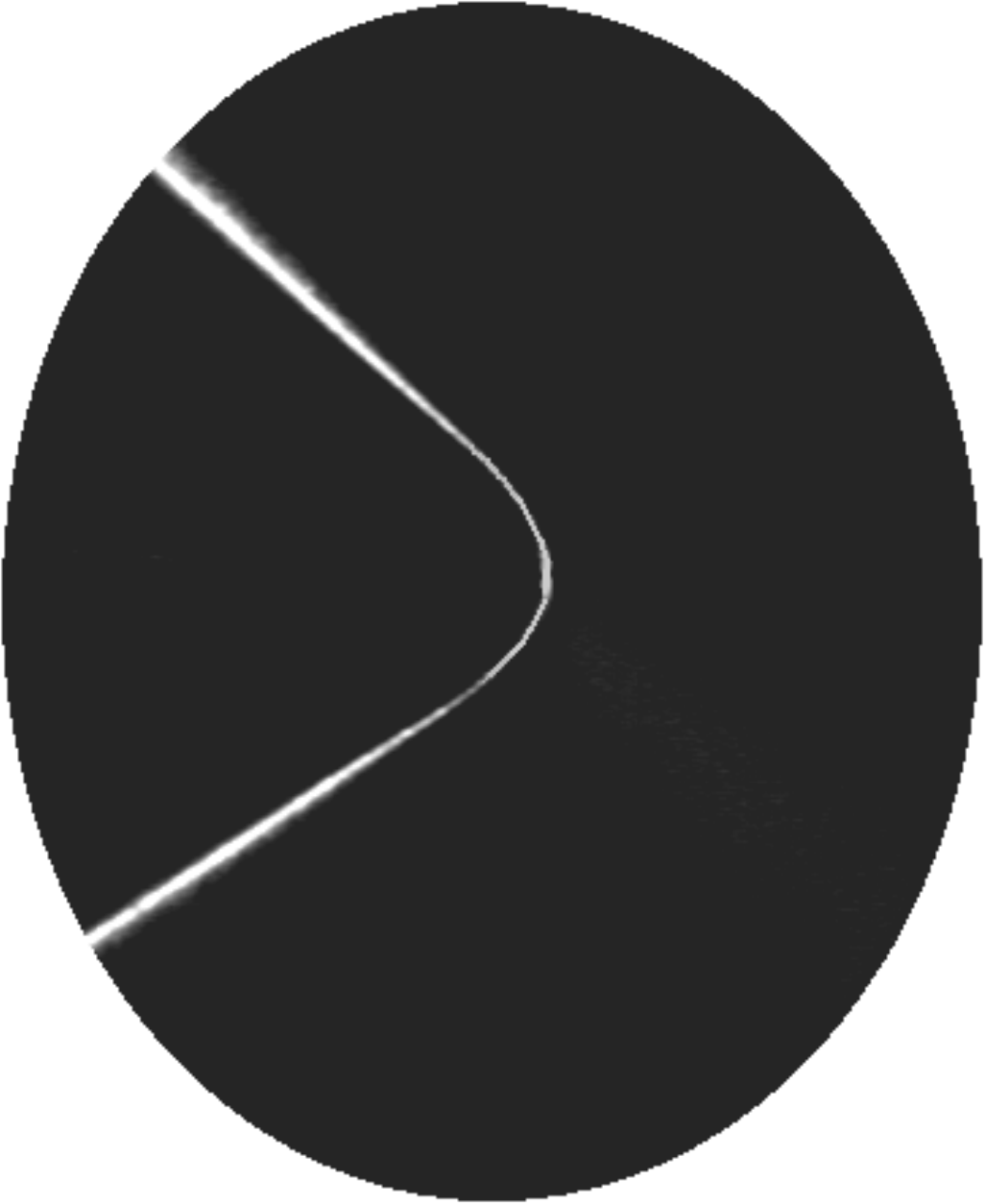}}
\caption{Top view of the d-cone at high deformation. The bright line which is no longer 
a parabola, but a hyperbola and one can notice a distance over which the wings of 
this line become linear. The deformation is about 7 mm. $R_f$=30mm}
\label{dc-hig}
\end{figure}

\begin{figure}
\centerline{\epsfxsize=\columnwidth \epsfysize=\columnwidth \epsfbox{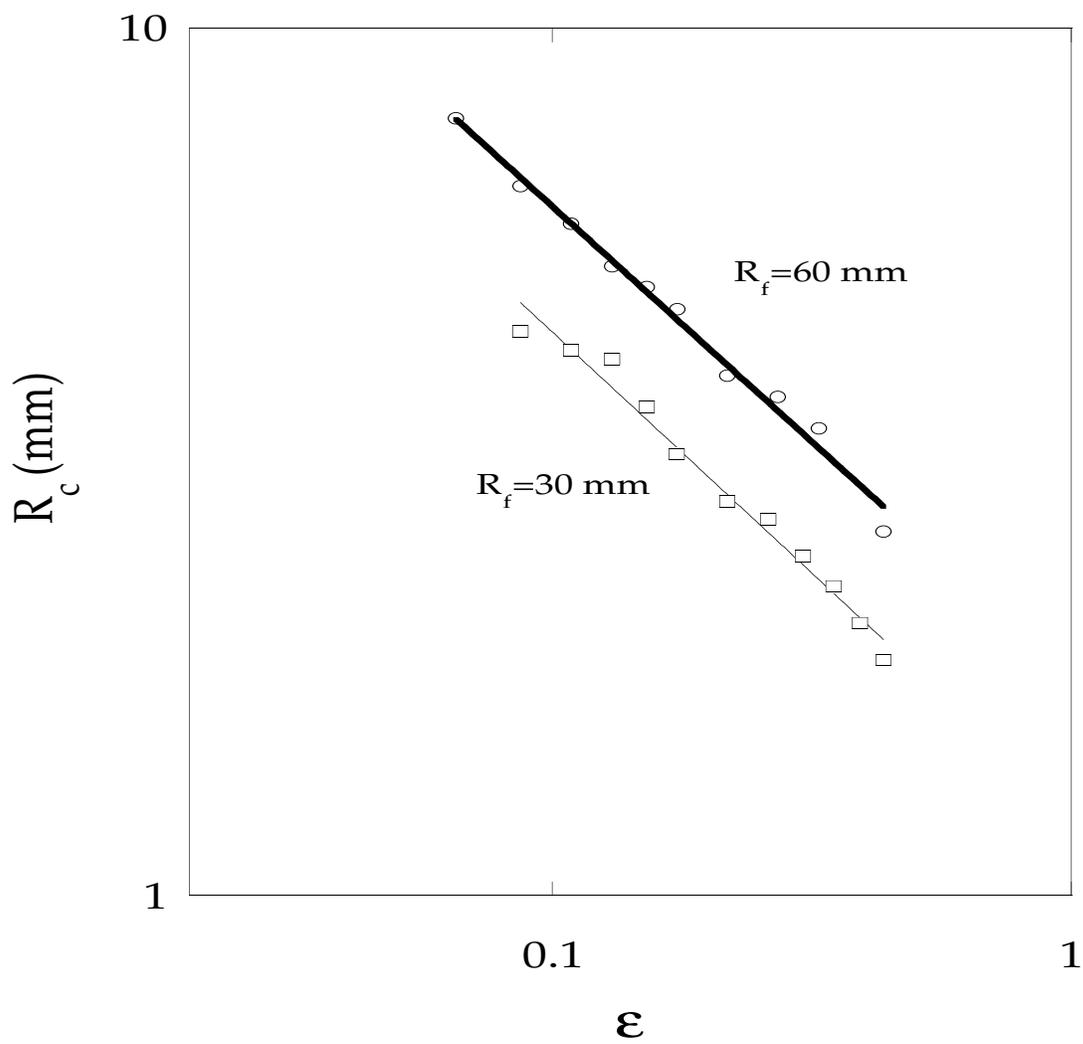}}
\caption{The radius of the crescent versus $\epsilon$ for large deformations. The
line are best fit to the power law $\epsilon ^{-1/2}$}
\label{r-hig}
\end{figure}

We have measured the curvature of the crescent at high deformation too. In this case
the ridge is a thin line and its shape is no longer a parabola. It has a shape of 
hyperbola, which wings are likely to be linear. To find the  crescent radius of 
curvature, we follow the same method for fitting
as above. A d-cone  at large deformation and highlighted from above 
is displayed in figure~\ref{dc-hig}.
From figure~\ref{dc-hig}, we notice that the bright line, separating the convex region and
the concave region, looks
like a hyperbola that define a sharp ridge near the core of the singularity, that is 
the region very close to the pushing tip, but its asymptotes (wings) 
 are straight lines \cite{wing}.
The radius of the crescent is measured by fitting the crescent to a polynomial in
the region close to the tip. Over a given distance the line does not belong to a hyperbola 
neither to a parabola. In figure~\ref{r-hig}, we show the radius of curvature of the 
crescent for $\epsilon >0.1$. The fitting procedure does not depends on the polynomial 
degree  we use to make the fitting. The lines in figure \ref{r-hig} are power 
laws with an exponent $-1/2$.

In the following we present a model based on a competition between pure bending and 
pure stretching but in the region that bounds the crescent.
\protect
\subsubsection{Scaling for high deformations}
In this section we will show that the power law can be found by considering that
the the concave region near the pushing tip is ,besides being stretched, bend as well. Also
we consider that the envagination is no longer moving downward, but the ridges are 
approaching each other. 
To do so let us write the bending energy 
\begin{equation}
E_{\text{ben}}=\kappa \int (\nabla ^2 \xi )^2 dS
\end{equation}
and the stretching energy for such a plate.
\begin{equation}
E_{\text{str}}=G\int (\nabla ^2 \chi )^2 dS
\end{equation}
Here $\chi$ is the Airy function, $\kappa$ is the bending rigidity, and $G$ is the 
stretching modulus and related to $\kappa$  for two-dimensional plates 
by $G \simeq\kappa /h^2$ where $h$ is the thickness. If we suppose that all the stretching
on the part that delimitates the concave part and the convex part (ridge) is due
to bending at the tip. The curvature $\nabla ^2 \xi$ can be then written as
$d/R_c^2$. When taking the derivative we suppose that all the curvature is due to
the deformation near the tip. We integrate the bending energy over the surface
$R_fR_c$. Hence, the bending energy is,
\begin{equation}
E_{\text{ben}} \sim \kappa (d/R_c^2)^2 R_f R_c \sim \kappa d^2 R_f R_c^{-3}
\end{equation}
For the Stretching energy we keep the same surface of integration. The strain
created by the decay of the curvature is due to the fact that the sheet is loaded
by a small angle $\epsilon$. The plate experiences a force $F$  when creating 
the deflection along the radius $R_f$, the plate has then a 
moment at the tip $FR_f$. For high deformations, the deflection not only 
moves downward but the two "wings" start to approach each other 
in the azimuthal direction which creates a characteristic strain 
$\gamma = (d^2/R_f^2)^2$. The stretching energy is then:
\begin{equation}
E_{\text{str}}\sim \kappa/h^2 (d^2/R_f^2)^2 R_f R_c \sim \kappa h^{-2}d^4 R_f ^{-3}R_c
\end{equation}
Minimizing $E_{\text{ben}} + E_{\text{str}}$ with respect to $R_c$, we find that:
\begin{equation}
R_c \sim R_f \sqrt {h / d}
\nonumber
\end{equation}
Knowing that $\epsilon =d/R_f$, 
\begin{equation}
R_c \sim \sqrt {R_f h /\epsilon}
\end{equation}

\begin{figure}
\centerline{\epsfxsize=\columnwidth \epsfysize=\columnwidth \epsfbox{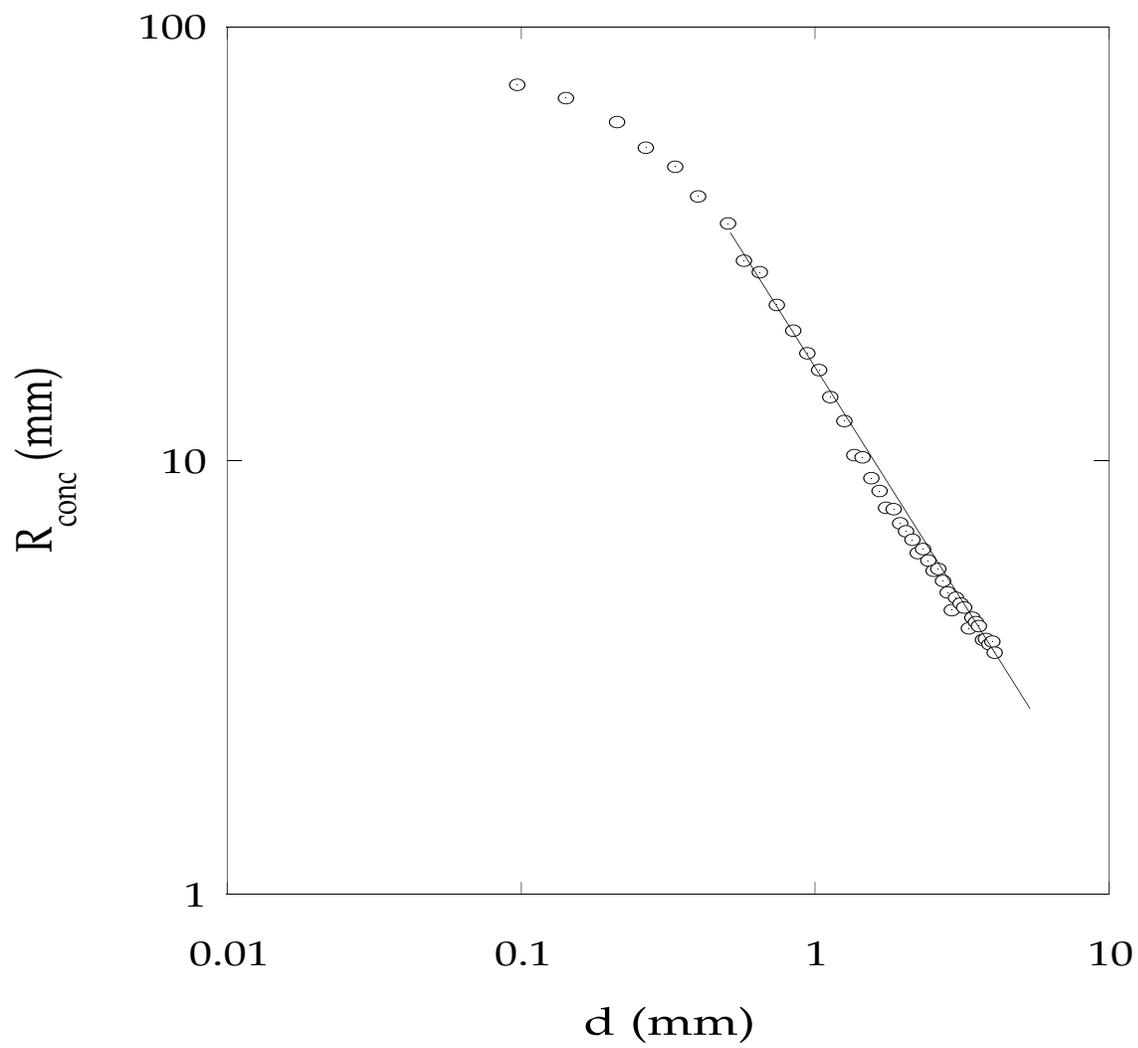}}
\caption{The local radius of curvature of the concave region close to the singularity. 
The line has  a slope close to -1.}
\label{Rconc}
\end{figure}

In \cite{maha}, it was shown that if we consider the tip of the d-cone as the core of a 
dislocation whose bending energy is logarithmic \cite{amar}, and if one considers 
two types of stretching (radial and azimuthal) one recovers the scaling of a d-cone 
obtained by squeezing a sheet in a cone of revolution i.e $R_c \sim 1/\epsilon$. 
This scaling is probably due to the fact that the only scaling in this particular problem, 
apart from the thickness, is the opening angle of the squeezing-cone .
In figure~\ref{Rconc} we show the radius of curvature of the d-cone concave part 
very close to the pushing tip.
For large deformations the radius scales like ${\epsilon}^{-1}$. It is possible that
the concave region behaves like a second cone which opening angle  is given by how
much it is squeezed in the cone defined by the convex region. 

\begin{figure}
\centerline{\epsfxsize=\columnwidth \epsfysize =\columnwidth \epsfbox{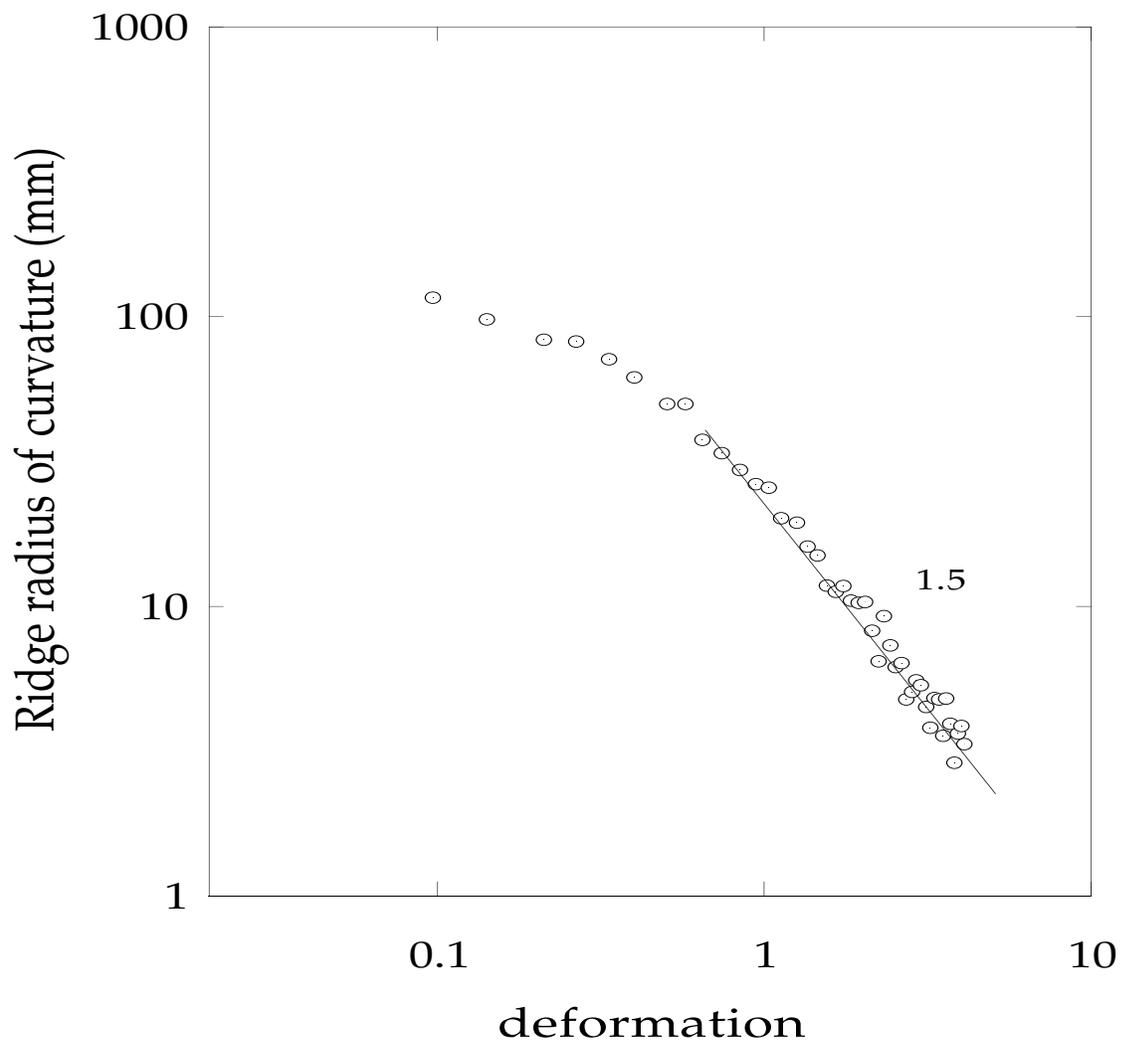}}
\caption{The radius of curvature of the ridge as a function of the deformation. The 
straight line has a slope - 1.5}
\label{ridge}
\end{figure}

One possible way to observe stress focusing is by comparing the growth of the curvature 
between the concave region and the ridge. Previously, we have shown that 
the cut-off radius $r_c$  in fig.~\ref{curv-str} decreases when increasing the
the deformation and this was due to stress focusing which gives rise to the crescent 
at the ridge. In figure~\ref{ridge}, we plot the radius of curvature as a function of
the deformation and the slope equivalent to the one in figure~\ref{Rconc} is equal to 1.5.
The fact that the radius of curvature at the ridge decreases faster than the radius at the
concave region is also a test of stress focusing inducing a curvature focusing.

In the next section we will show how from the profiles one can also observe 
stress focusing by measuring the reaction force at the ridge and at the concave part.

\protect
\section{Singularity Energy and Force measurements}
The crumpled paper is similar to the discovery due to Laplace and best known as
the ``Plateau Problem'' which consists of the determination of minimal 
surface given its border (soap film), whereas in the crumpling problem, instead, 
the volume is kept fixed ; that is what one does when making a ball by crumpling a piece of 
paper, the stress in this case is distributed on the regions where the sheet is ``overfolded''.
The d-cone problem is much simpler than the real crumpled paper. It consists as 
discussed above of fixing a border, that is the frame radius (fig.~\ref{disc}) and 
squeezing a sheet into it. The sheet is then buckled and a sharp part is created at the 
pushing point. The force applied on the center of the plate is responsible of the creation 
of the torque that causes the plate to deflect and gives rise to the ``d-cone''.  In this 
section we discuss the results of the response of the plate to the external load at its center. 
\protect
\subsection{Torque from profiles}
\begin{figure}
 \vskip -0.3in
 \centerline{\epsfxsize=\columnwidth \epsfysize=\columnwidth \epsfbox{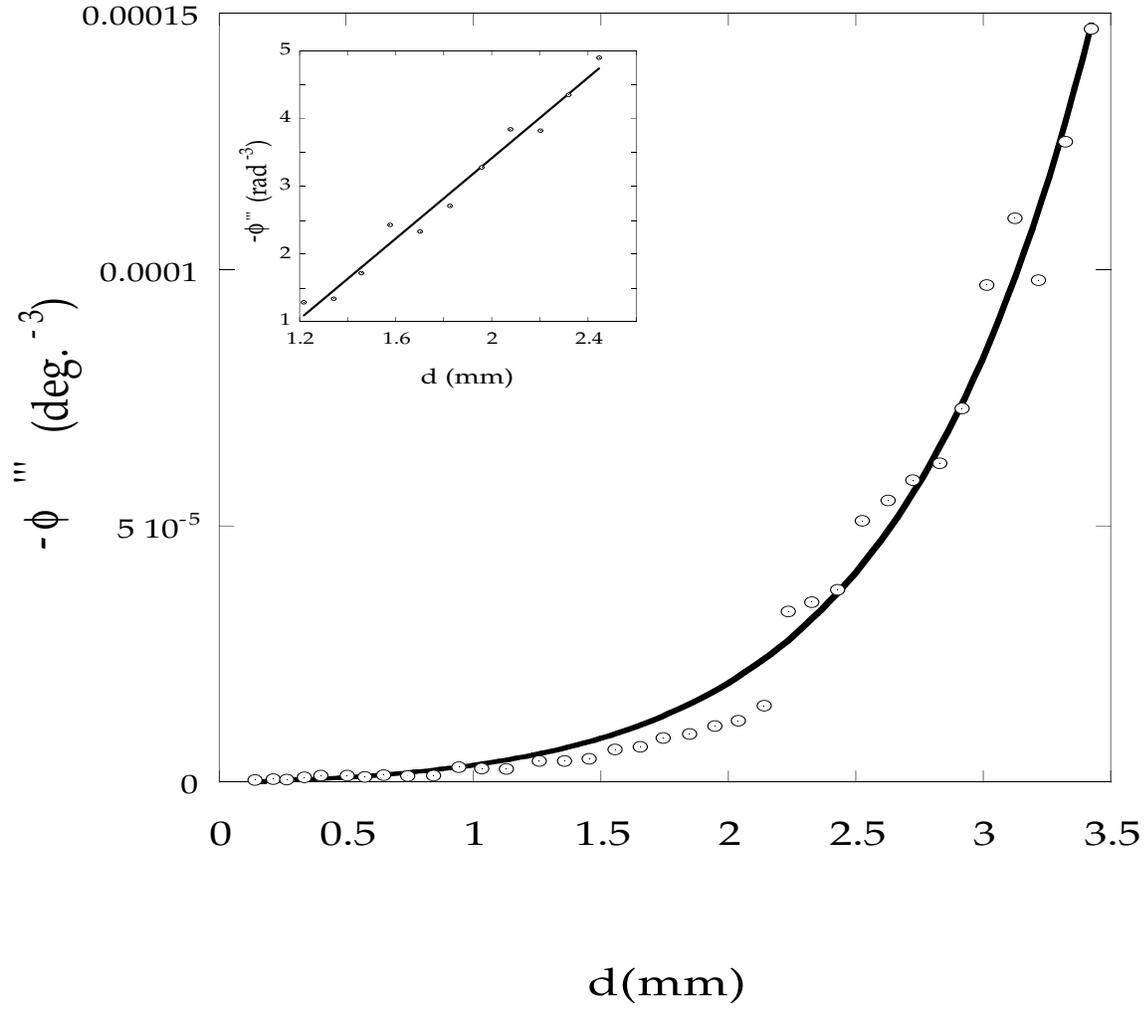}}
 \caption{The third derivative of the profile ($\phi$) at the ridge as versus the 
 deformation and at a distance of 3 mm from the tip. Inset: the third derivative 
 at a distance of 6 mm from the d-cone tip. }
 \label{psi}
 \end{figure}

In this section we will determine the reaction forces  experienced by the 
plate at the borders as a result to the  external load (F). The resultant of these 
forces is equal to the external loading ; we will determine the force at the ridge where 
most of the stress is concentrated
(fig.~\ref{phitheta}). It is well known in classical mechanics 
that the force acting in a certain direction  is equal to the derivative of the energy 
with respect to the coordinate in this direction. In our case, the reaction force 
experienced by the plate at the border where it is  deflected, is defined by the 
derivative of the energy with respect to the displacement $\zeta$. The plate resists
bending by a reaction force: 
 \begin{equation}
 F_{\text{reac.}}=\frac{\partial E_b}{\partial \zeta}\sim \frac{1}{\zeta '}\frac{\partial }{\partial \theta} \int 
 \Bigl(\frac{\partial^2 \zeta}{\partial \theta ^2}\Bigr) ^2 \,d\theta\nonumber
 \end{equation}
 Integrating by part we find:
 \begin{eqnarray}
 F_{\text{reac.}}&\sim& - \frac{1}{\zeta '}\frac{d}{d\theta} \int \zeta ' \, \zeta ''' \,d\theta \\
 &\sim&  -\zeta ''' =-r\phi ''' =-r\frac{{\partial }^3 \phi}{{\partial \theta ^3}}
 \end{eqnarray}
 In figure~\ref{psi}
 we display the third derivative of the profile at the ridge. To measure the reaction force
 we derive the profile twice versus the angle $\theta$ and we calculate the slope of 
 the straight line where the curvature $ \sim \zeta ''$ varies. The
 force is supposed to increase linearly for small deformations when the regime is still 
 elastic. From figure~\ref{psi}, $\phi ''' $ blows up exponentially, we believe that this is
 due to the plastic transition. 
 However, far away from the
 singularity the force increases linearly even for large deformations. The fit in 
 figure~\ref{psi} is of the form ($d e^{d/a}-1$), whereas the fit in the inset is 
 a linear fit.
 
\begin{figure}
\centerline{\epsfxsize=\columnwidth \epsfysize=\columnwidth \epsfbox{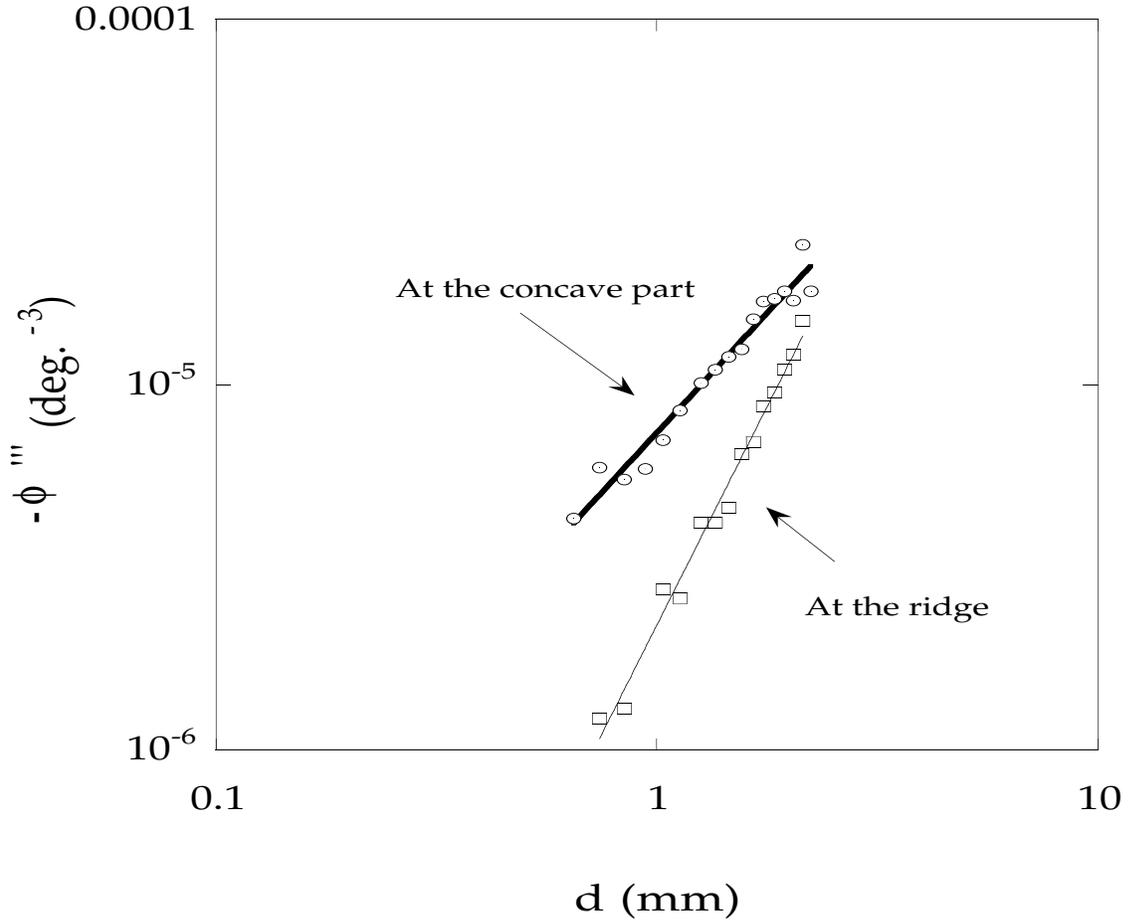}}
\caption{The reaction force at the ridge and at the concave part vs. the deformation. The 
slope of the line giving the force at the ridge is twice larger than the one corresponding to
the concave part. The data are taken at 3 mm from the tip.}
\label{p1p2}
\end{figure}

\begin{figure}
\centerline{\epsfxsize=\columnwidth \epsfysize=\columnwidth \epsfbox{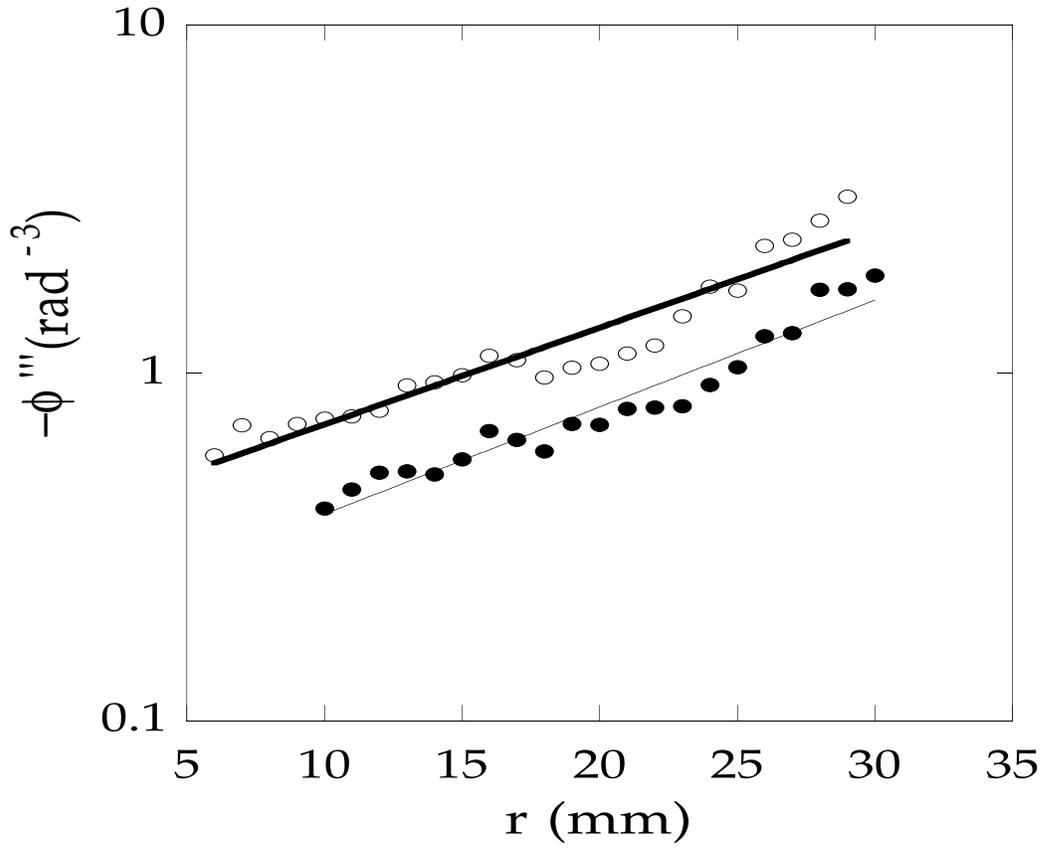}}
\caption{The reaction force at the ridge vs $r$. The upper curve corresponds to d=1.9 mm
and the lower curve corresponds to d=1.2 mm.}
\label{phir}
\end{figure}

 The exponential behavior of the force, is also a measure of the force focusing which 
 creates the crescent near the tip. The exponential behavior was observed in the
 behaviour of the curvature for high deformations (fig.~\ref{curv-str}) and was 
 explained as a consequence of geometry-induced stretching. To satisfy energy 
 considerations and to verify scaling considerations, this curvature was found to decay 
 exponentially \cite{lobko1}. The reaction force ($\sim |\phi '''|$) far from the singularity
 is linear over the same range of deformation (Inset of fig.~\ref{psi}).
Two regions experience reaction forces and then torque which gives rise to the
invagination. These two region are: the ridge and the concave part. It can be easily 
observed that the ridge experiences more stress than the concave part. In the 
fig.~\ref{p1p2} we display $ |\phi '''|$ at the ridge and at the concave region.
The slope of the fitting line in the case of the reaction force at the ridge is larger than the
one corresponding to the concave part.
It is clear then that the stress due to the reaction force increases 
faster at the ridge than at the concave part. The stress is focused at the ridges where the
crescent takes place. 
As the plate is more deflected far away from the singularity than closer to it, we have 
measured the reaction force as a function of the distance from the singularity for two 
different  deformations. Figure~\ref{phir}, displays $-\phi '''$ versus the distance $r$.

We would have a linear behavior for the reaction force if only bending is present, but as 
the plate may be stretched, we have a non-linear increase of the reaction force. However 
the behavior of the reaction force seems to be the same for the two distances.
In the following 
paragraph, we discuss a measure of the force experienced by the plate at its tip.
 
\protect
\subsection{Direct Force measurements}
\begin{figure}
\centerline{\epsfxsize=\columnwidth \epsfysize =\columnwidth \epsfbox{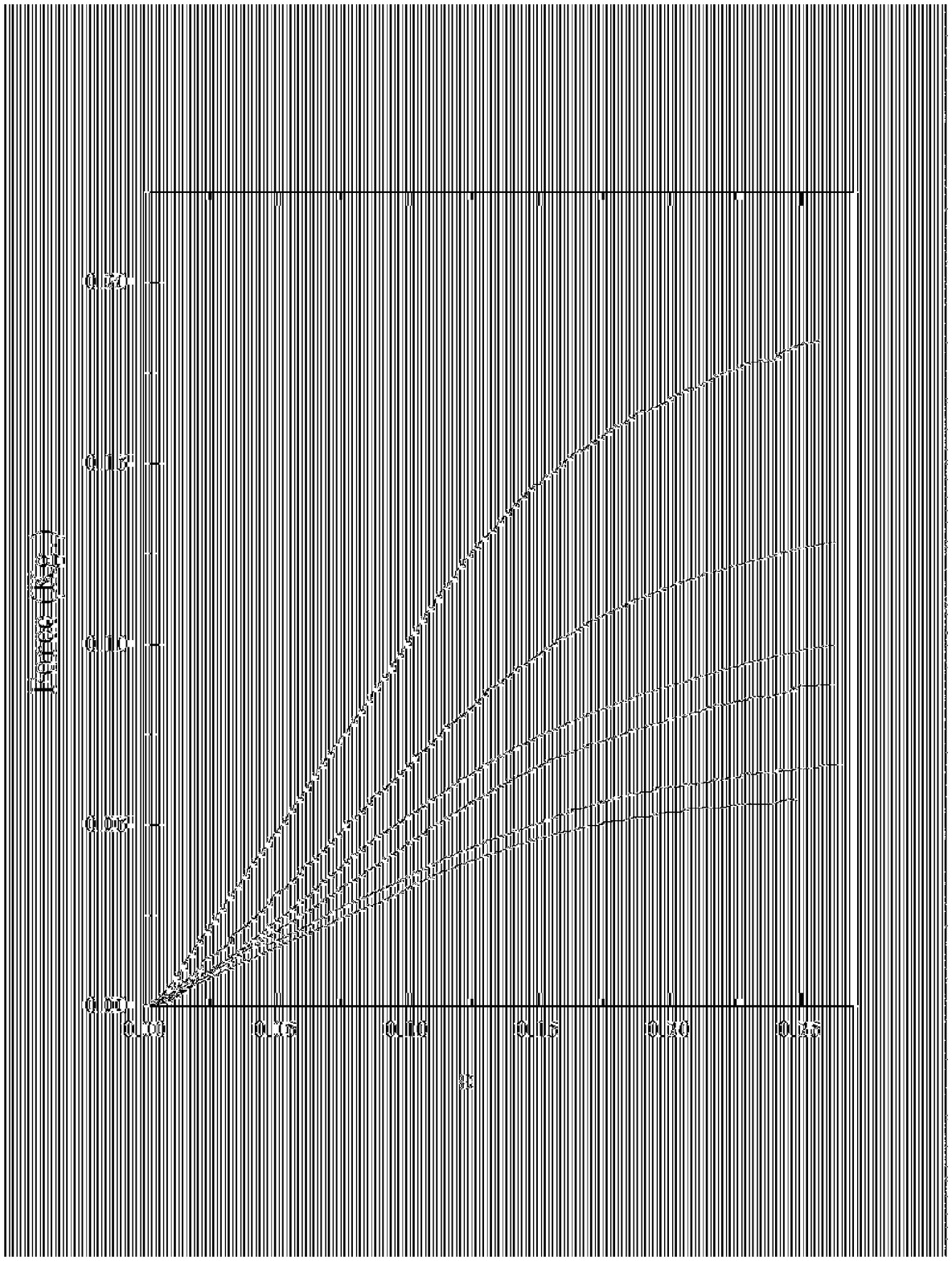}}
\caption{Force vs. $\epsilon$ for different frame radius for transparencies.
 The saturation force increases when the frame radius decreases. For stailess steal, the
 crossover is an order of magnitude lower. For copper the crossover is 0.05.}
\label{force}
\end{figure}

When the plate is pushed and at for low $\epsilon$, a deflection appears and neither a 
singularity nor a sharp ridge
is observed: This regime is elastic. The force exerted by the plate 
is linear in the deformation. As we increase the load, the line at the
ridge become sharper and the force is no longer linear: the regime is plastic. In 
figure~\ref{force}, we display the force versus $\epsilon$. It is clear that the opening 
angle defined by $\epsilon$ at
which the force saturates is the same for the different frame radius. However, the 
maximum force at saturation is large for smaller frames. Also, it is worth noting that
the force changes its slope around $\epsilon \simeq 0.1$ for transparencies, 
this correspond to the same value where we have observed a crossover between 
the $-1/3$ and the $-1/2 $ scaling of the crescent radius versus $\epsilon$. 
It is possible, as pointed out earlier, that the ``folding'' mode has changed from a regim 
to another. For copper the crossover value is equal to 0.05, and for stainless steal, it is an order of 
magnitude smaller than for transparencies.

\begin{figure}
\centerline{\epsfxsize=\columnwidth \epsfysize=\columnwidth 
\epsfbox{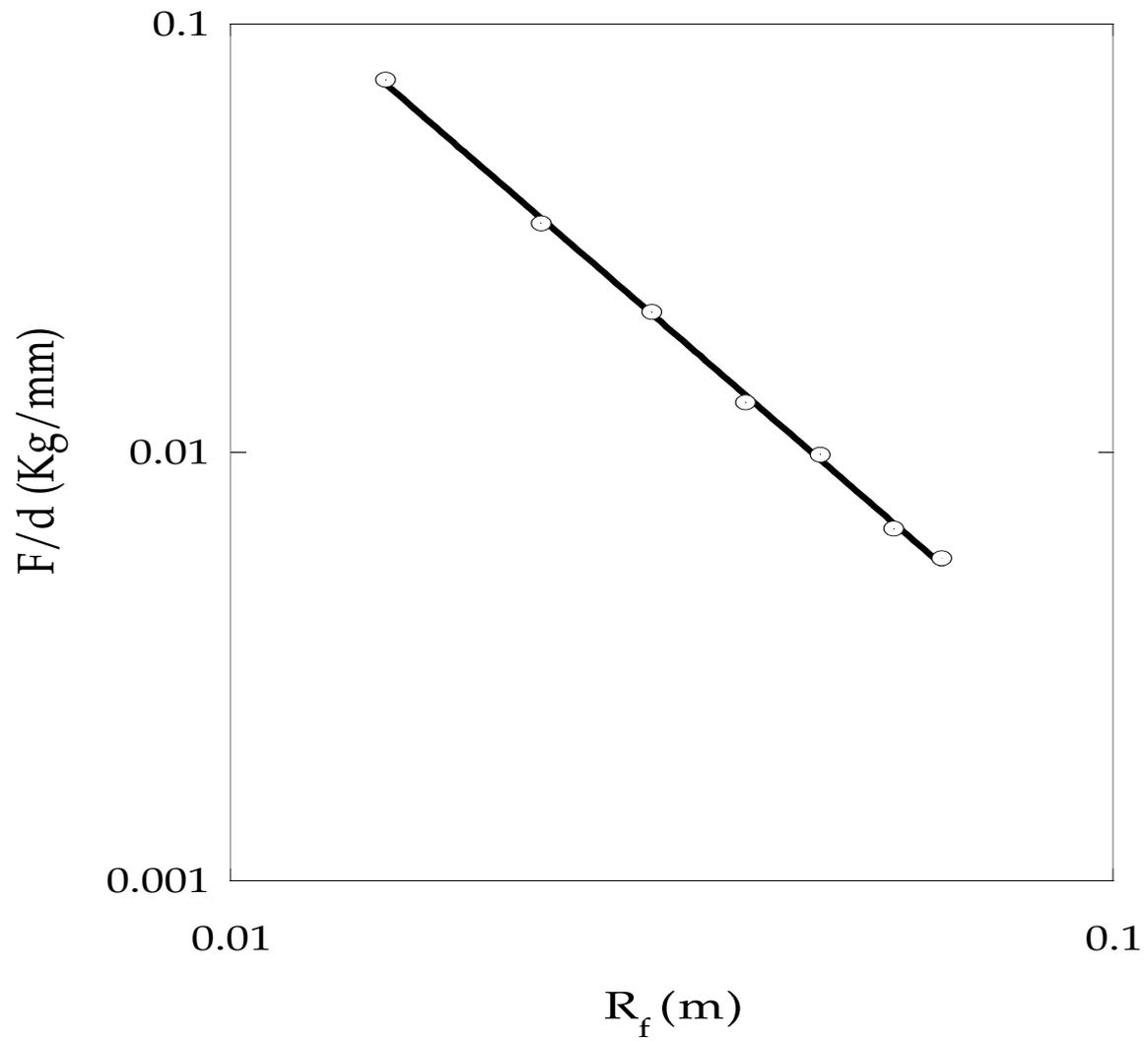}}
\caption{The slope of the force in the elastic regime vs. $R_f$. The slope of the 
fitting line is close to 2.}
\label{scal}
\end{figure}

In figure~\ref{force} the maximum force at which the plate saturates scales with 
the frame radius like $F_{\text{sat}} \sim R_f^{-0.77}$. For low deformations the force 
is linear with the deformation where only bending is dominant. 
If we assume that in the elastic regime, the work necessary to load the plate by a distance
$d$ is ($F\times d$) and is equal to the bending energy which is proportional to 
$d^2/R_f^2$ (to be integrated over 
the plate surface $R_f^2$), we find that $F \sim d/R_f^2 $. This result is in agreement 
with the behavior of the reaction force in figure~\ref{psi}. In figure~\ref{scal}, we plot
the slope $F/d$ of the linear part of the force versus the frame radius. The slope of the 
fitting line is close to 2. From this scaling, the force goes like $\epsilon /R_f$.
Now we are able to measure the energy necessary to create the crescent line. In the 
following we show how we measure this {\em singularity energy}.
\protect
\subsection{Singularity Energy}
\begin{figure}
\centerline{\epsfxsize=\columnwidth \epsfysize =\columnwidth 
\epsfbox{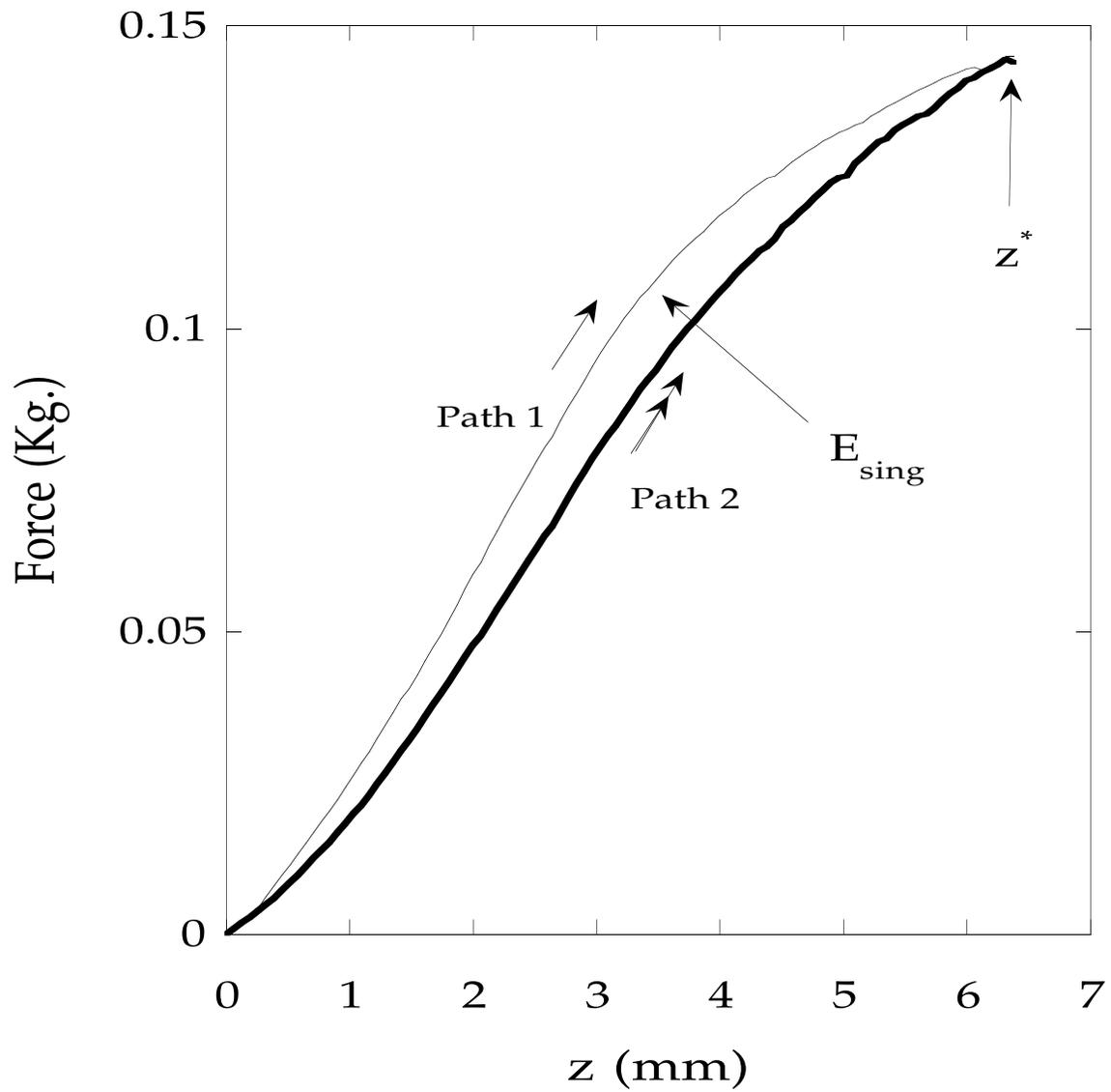}}
\caption{The method how the energy of the singularity is measured. Path1 
and Path2 are the ``crumpling'' of a perfectly flat plate and the reloading of
the already crumpled plate. The area between the thick line (path2) and 
the thin line (path1) is the energy of the singularity}
\label{twopath}
\end{figure}

When we bend a thin plate, well before the plastic regime is reached, and 
release it, it recovers its shape it has before bending. But, when we bend 
the plate till a point where the internal face feels compression and the 
external one feels stretching, the plate does not recover its shape. 
Although, this cannot be observed from the appearance of a scar,
in force measurements, the load necessary to bend the plate to the 
same point, is different and smaller because the plate is already ``weakened''. In order 
to measure the singularity energy, we measure the force up to a 
deformation $z^*$, and we release the plate, and we measure the force another time 
by reloading up to the same point $z^*$. The resultant area
between the line is the singularity energy, which corresponds to the energy 
dissipated in creating the scar region. If we load the plate a third time, the force follows 
the same thick line in figure~\ref{twopath}. 
In figure~\ref{twopath}, we show the two paths and the area between them as the 
energy of the singularity. 

\begin{figure}
\centerline{\epsfxsize=\columnwidth \epsfysize=\columnwidth \epsfbox{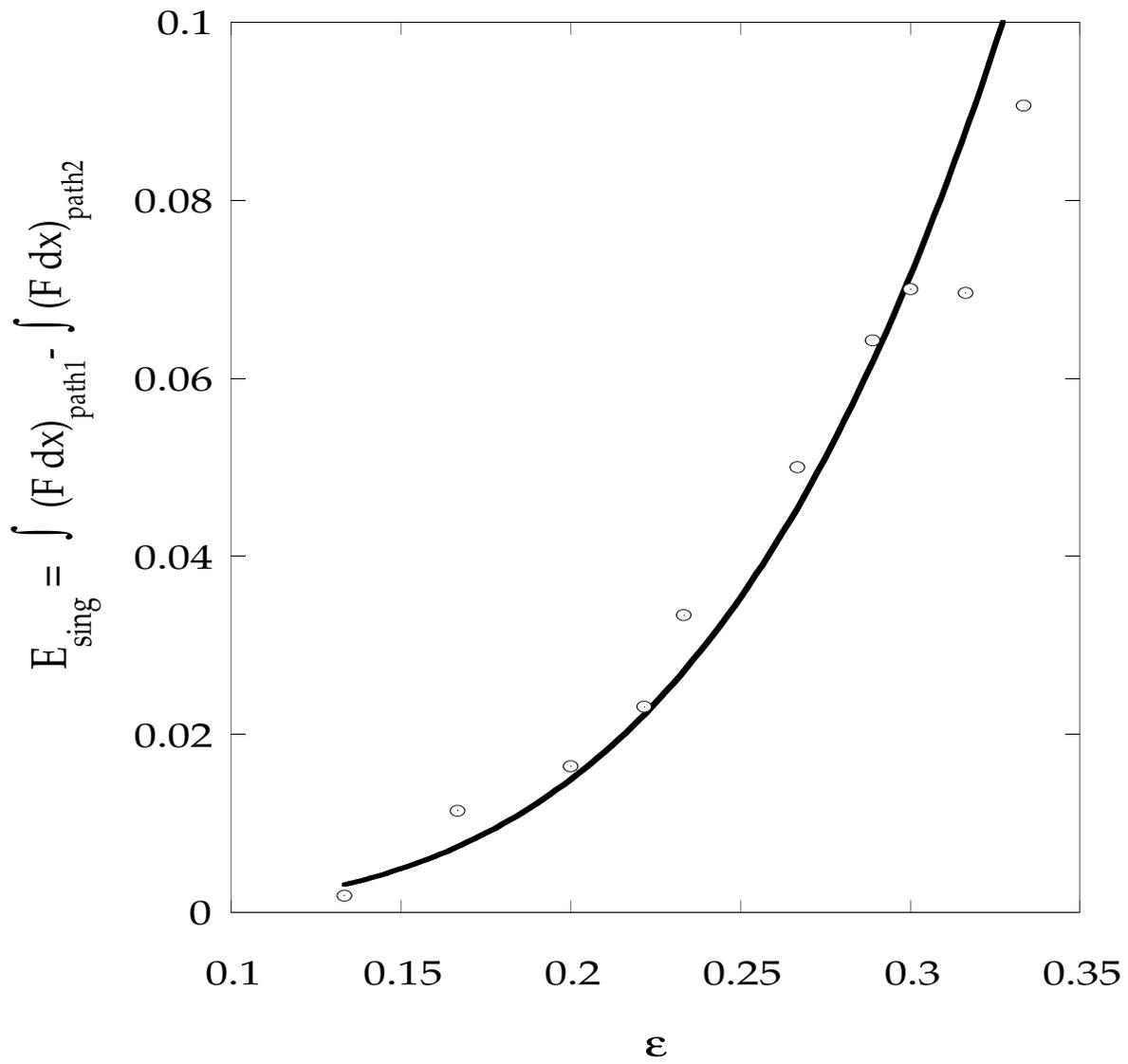}}
\caption{The singularity energy, which is the energy necessary to form the scar. 
The line is the best fit to a power law $\epsilon^4$}
\label{sing}
\end{figure}

The singularity energy
is seen as the energy dissipated during the scar formation. If $z$ is the
displacement, then the singularity energy is given by:
\begin{equation}
E_{\text{sing}}\simeq\int_{0}^{z^*}(F_1\:dz)_{\text{path1}}-\int_{0}^{z^*} (F_2\:dz)_{\text{path2}}
\end{equation}
Where path1, paht2, are the first run and the second after the scar has been formed. 
$F_1$ 
and $F_2$ are, the force on a new plate and on an already deformed plate respectively.
After the second run, the force always follows the path2 as long as the deformation 
does not exceed $z^* $. To measure the friction of the border on the plate, 
we ``unload'' the plate and measure the force during the ``unloading'', and for the same 
deformation, the force shows a sharp drop. This drop in
the force is due to friction and it is not taken into account, as it is eliminated when 
calculating the surface separating the two loads.
Figure~\ref{sing}, displays the Singularity energy as a function of $\epsilon$. Each 
point in the $x$ axis, corresponds to the point $z^*$ at which the plate was buckled.
As defined before $\epsilon = z/R_f=z^* /R_f$. The line is a power law ${\epsilon}^4$

In \cite{amar}, it was shown that the stress in the inner region, that is the 
region where the non linear effects are dominant, is proportional to 
$\epsilon ^2$. In their case, the total energy was reduced to the stretching energy : 
$\int (\partial ^2 \chi / \partial x_i \partial x_j)^2 dS \simeq \int {\sigma}_{ij} ^2 dS$. 
They have neglected the bending energy as it is proportional to $h^3$, and any other 
length scale in the problem is much bigger than $h$. For large deformations, 
this assumption is valid, because the frame radius is the most important length scale in
the problem as it works to squeeze the plate. 
Although this energy is the energy corresponding to the work done to deform 
the plate and stretch the plate to make the crescent, it can be seen as the ``d-con energy''.
When a d-cone is formed by forming a scar the plate has a quenched form and it does not 
recover its shape when one releases it. However, if one isolates the scar region, by 
cutting the a circular region around the singularity, the outer band becomes flat as if it was
purely bent. This region, may correspond to the ``wing region'' where the 
non linear effects are just felt but the plate geometry is not affected in an irreversible
way.
The scar region gives the plate the conical form, it is then legitimate 
to call the singularity energy a ``d-cone energy'' because all the energy dissipated 
is concentrated in the scar region.
\protect
\section{Discussion}
In this article we explored properties of the conical singularity in a thin 
elastic sheet known as the
developable cone that may be relevant to a description of crumpled elastic 
membranes. Profiles of the surface has been studied from which we derived the 
relation between the maximum deflection and the deformation imposed on the plate
and the anisotropy, which is a geometrical anomaly for real sheets ($h\ne 0$) related 
to the rejection of the singularity far from the plate to minimize energy and to keep 
the plate smooth at small deformations. The aperture angle of the d-cone was 
measured and was found universal and depends only on the geometry of the frame. 
A simple model  based on the minimization of the curvature energy for an isometric 
deformation was sufficient to describe the region outside the singularity. 
Also the shape of the d-cone obtained from profilometry measurements was reproduced 
using a simple geometrical model. The curvature
at the ridge and at the concave part were both measured, and a stress
focusing was observed as a fast decrease of the radius of curvature
at the ridge in comparison to the one at the concave region. From the profiles, 
it was possible to quantify the reaction force and describe how the stress is focused 
at the ridge where the
scar, in a form of a crescent, appears. This crescent has a curvature that scales with the
deformation experienced by the plate. Two regimes were found. At small deformations 
the crescent has a parabolic form whose radius varies slowly with the deformation and  
the singularity size span a region of the order of the frame radius. At higher deformations, 
the crescent is 
squeezed and the crescent is no  longer a parabola, and its radius varies faster with the
deformation; the singularity is confined to a smaller region around the tip. From load
measurements, the scaling of the force in the elastic regime versus the deformation and
the frame radius was found. The two different scalings of the crescent radius versus 
the deformation arise from the fact that at small deformations and when the force 
is linear versus the deformation, the plate is bent in the vertical direction and no sharp 
ridges is observed while the aperture angle $2\theta_0$ remains constant. 
Beyond the crossover where $\epsilon \simeq 0.1$ (for transparencies) and 
where the force changes its slope,the mechanisme is no longer the same: the two ridges
approach each other and 
 the crescent is itself folded in the azimuthal
direction and the aperture  angle decreases. Also, it was also possible to measure
the energy needed to create the scar, what we called the {\em singularity energy}.
This energy was measured as the energy dissipated in the plate to form the 
scar. As the scar region is more rigid than the rest of the plate due to plastic effects, 
this region sustains the rest of the plate, the energy of the d-cone is then concentrated 
within this region. The stress focusing inducing a curvature focusing 
and an increase in the singularity energy is similar to the defects-induced 
dislocation in liquid crystal, where the defect when squeezed in a smaller region
its spatial extension is decreased and the
curvature of the molecular planes increases  \cite{oswald}. In fact, if we look at a 
smectic layer in the vicinity of a core of parabolic domain, we can notice that
this layer defines an object similar to a d-cone. In this experiment the stress focusing 
induces a strain localizarion near the singularity. From this study, it is worthy to note 
that the d-cone problem is very similar to the dislocation problem \cite{maha1} and will be
the subject of a futur work \cite{chaieb5}. Apart 
from the analogy between the logarithmic divergence of the curvature energy, 
the deflection resembles a Orowin's analogy of the motion of a snake or a carpet 
\cite{maha}. The strength of the dislocation in this case is measured by the maximum 
deflection. Although crumpled vesicles has been observed \cite{ben}, no systematic 
local study of the surface of a crumpled vesicle has been performed. 
Profilometry using laser beam or magnetic beads on the surface of a crumpled vesicle 
can complement freeze fracture microscopy experiments usually used to probe vesicles 
in suspensions. It is known that the creases nucleation in a buckled plate is subcritical 
and the formation of the singularity that bounds a crease is sudden, it is then of great 
importance to study 
topological properties of such singularities at subcriticality \cite{chaieb1}. Also, in real 
crumpled sheets, singularities may interact giving rise to a rich behavior as the one 
encountered in the physics of defects.

\protect
\section*{acknowledgments}
We thank Jean-Christophe G\'eminard for his valuable collaboration in sections IV and 
Va, Vb and Vc. We thank E. Cerda and L. Mahadevan for sharing us their points
of views.
This work was financed in part by the DICyT of the University of Santiago Grant 
No. 049631CH, and by a {\em Catedra Presidencial en Ciencias}.

\bibliographystyle{unsrt}

\begin{thebibliography}{10}

\bibitem{ben}
M.~Mutz, D.~Bensimon, and M.J. Brienne.
\newblock Phys. Rev. Lett. {\bf 67},923 (1991).

\bibitem{kantor}
Y.~Kantor, M.~Kardar and D.R.~Nelson,
\newblock Phys. Rev. Lett. {\bf 57}, 791 (1986); {\em Statistical Mechanics
of Membranes and Surfaces}, edited by D.~Nelson, T.~Piran, and S.~Weinberg,
(World Scientific, Singapore, 1989).
\bibitem{chaieb}
D.~R.~Nelson and L.~Peliti,
\newblock J. Phys. (Paris) {\bf 48} 1085 (1987);
\newblock M.~Paczuski, M.~Kardar and D.~R.~Nelson,
\newblock  Phys. Rev. Lett. {\bf 60} 2638 (1988);
\newblock D.~Morse, T.~Lubensky and G.~Grest,
\newblock  Phys. Rev. A {\bf 45} R2151 (1992);
\newblock D.~Bensimon, D.~Mukamel and L.~Peliti,
\newblock Europhys. Lett. {\bf 18} 269 (1992);
\newblock X. Wen, C. W. Garland, T. Hwa, M, Kardar, E, Kokufuta, Y. Li,
\newblock M. Orkisz and T. Tanaka, Nature {\bf 355} 426 (1992);
\newblock R.~Attal, S.~Cha\"{\i}eb, and D.~Bensimon.
\newblock Phys. Rev. E {\bf 48} 2232 (1993);
\newblock Y. Park and C, Kwon, 
\newblock Phys. Rev. E {\bf 54} 3032 (1996).
\bibitem{spec}
M.~S.~Spector, E.~Naranjo and J.~A.~Zasadzinski,
\newblock Phys. Rev. Lett. {\bf 73} 2867 (1994); 
\newblock  R.~R.~Chianelli, E.~B.~Prestige, T. ~A.~Pecoraro and J.~P.~DeNeufville,
\newblock  Science {\bf 203} 1105 (1979). 
\bibitem{falvo}
M.~R.~Falvo, G.~J. ~Clary, R.~M.~Taylor~II, V. ~Chi, F.~P.~Brooks~Jr, 
S.~Washburn and R. ~Superfine,
\newblock Nature {\bf 389} 582 (1997);
\newblock M.~M.~Treacy, T.~W.~Ebbesen and J.~M.~Gibson,
\newblock Nature {\bf 381} 678 (1996);
\newblock B.~I.~Yakobson, C.~J.~Brabec and J.~Bernholc,
\newblock Phys. Rev. Lett. {\bf 76} 2511 (1996).
\bibitem{deser}
S.~Deser, R.~Jackiw and G.~'t Hooft,
\newblock Ann. of Phys. {\bf 152}, 220 (1984).

\bibitem{pome}
Y.~Pomeau,
\newblock C. R. Acad. Sci. I, Math. {\bf 320}, 975 (1995); 

\bibitem{lobko} 
A.~Lobkovsky, S.~Gentges, 
Hao.~Li, D.~Morse, and T.A. Witten,
\newblock  Science {\bf 270}, 1482 (1995).

\bibitem{amar}M.~Ben Amar and Y.~Pomeau, 
\newblock  Proc. R. Soc. London A, {\bf 453},729, (1996).


\bibitem{witt}
T.A. Witten and H.~Li, 
\newblock Europhys. Lett. {\bf 23},  51 (1993); 
\bibitem{kramer}E. M. Kramer and T.A. Witten,
\newblock Phys. Rev. Lett. {\bf 78}, 1303 (1997).

\bibitem{lobko1} A.~Lobkovsky, 
\newblock  Phys. Rev. E {\bf 53}, 3750 (1996); 

\bibitem{lobko2} A,~E.~Lobkovsky and T.~A.~Witten
\newblock Phys. Rev. E {\bf 55}, 1577 (1997)

\bibitem{chaieb1} S.~Cha\"{\i}eb and F. Melo,
\newblock Phys. Rev. E  {\bf 56}, 4736 (1997).

\bibitem{chaieb2} S.~Cha\"{\i}eb, F.~Melo and J.~C.~G\'eminard,
\newblock  Phys. Rev. Lett. {\bf 80}, 2354 (1998)

\bibitem{maha} E. Cerda and L. Mahadevan, Phys. Rev. Lett. {\bf 80}, 2358 (1998).
\bibitem{chaieb3} S. Cha\"{\i}eb and J.C. G\'eminard (in preparation).
\bibitem{chaieb4} S. Cha\"{\i}eb and F. Melo in Instabilities and Nonequilibrium 
Structures VI, Tirapegui and Zeller Eds (1997).
\bibitem{maha1} L. Mahadevan (private communication)
\bibitem{chaieb5} S. Cha\"{\i}eb, E. Cerda, F. Melo and L. Mahadevan (in preparation)
\bibitem{landau}
L.~Landau and E.~Lifchitz,
\newblock {\em Th\'eorie de l'\'elasticit\'e}.
\newblock  (Mir, Moscow,1967)  page 87.

\bibitem{wing} The term (wing) is introduced in \cite{lobko2} to describe the region 
where the strain persists although all the energy is concentrated near the
singularity.
\bibitem{oswald}
P. Oswald (private communication)

\end{thebibliography}



\end{document}